\documentclass[showpacs,preprintnumbers,prd,showkeys,aps]{revtex4}
\usepackage{graphics}
\usepackage{eurosym}
\usepackage{amsfonts}
\usepackage{amssymb}
\usepackage{amsmath}
\usepackage{graphicx}
\usepackage[font={footnotesize,it}]{caption}
\usepackage{amsmath,array}

\begin{document}

\title{Black hole radiation of massive spin-2 particles in (3+1) dimensions}
\author{I. Sakalli}
\email{izzet.sakalli@emu.edu.tr}
\author{A. \"{O}vg\"{u}n}
\email{ali.ovgun@emu.edu.tr}
\affiliation{Physics Department, Eastern Mediterranean University, Famagusta, Northern
Cyprus, Mersin 10, Turkey}
\date{\today }

\begin{abstract}
This paper is devoted to the study of radiation of massive spin-2 boson
(graviton with a nonzero mass) through the event horizon of a generic static
and spherically symmetric black hole in (3+1) dimensions. To this end, we
consider the problem in the framework of quantum tunneling phenomenon. We
evaluate the tunneling rate of the massive gravitons by applying the
semiclassical WKB approximation to the Fierz-Pauli equation. The temperature
of the radiation is obtained with the aid of the Boltzmann expression. Our
findings are in good agreement with the existing Hawking radiation studies
in the current literature.
\end{abstract}

\maketitle

\section{Introduction}

Hawking \cite{H0,H1,H2,H3} theoretically proved the existence of black hole
radiation: a thermal (black-body) emission. It is named after him as the
Hawking radiation (HR). This radiation causes in fact from the steady
conversion of quantum vacuum fluctuations into pairs of particles, one of
which escaping at spatial infinity while the other is caught by the black
hole at the event horizon. This radiation reduces the mass of black holes
and is therefore also known as black hole evaporation. Furthermore, each
evaporation of different black holes possesses a characteristic temperature
which is the so-called Hawking temperature. Since Hawking's those seminal
papers, many physicists have been studying on the different derivations of
the HR for the numerous spacetime models including black holes, wormholes,
black strings etc. by using different kind of quantum particles
(spin-0,1/2,1,3/2...). For example, a reader may refer to \cite%
{Wil1,Wil2,Wil3,Pad1,Pad2,Pad3,Unruh,Pad4,Mann1,
Mann2,Jing,Majhi,Vanz,kraus1, kraus2,perkih2,perkih3,ang,
kruglov1,sh2,xiang,kruglov2,ak1,ia1,
ia2,ia3,ia4,iak1,ia5,ak2,gohar1,Akh1,Akh2,pas0,sakhal,gup05}. 

One of the most well-known methods in the computation of the HR is to use
the WKB\ approximation \cite{KG} in the complex path integration method of
Srinivasan et al. \cite{Pad1,Pad4}. In this approach, one usually considers
energetic quantum particles which penetrate the black hole through its
horizon. Such quanta have classically forbidden (+) / acceptable (-)
trajectories whose action ($S_{\pm }$) is suitable for the WKB approximation
method. The differential equation of the action can be obtained by
substituting the proper wave function $\mathit{\Phi}\sim e^{\frac{i}{\hbar }%
S(t,r,\theta,\phi)}$ into the wave equation of a considered curved
spacetime. Then, the Hamilton-Jacobi method \cite{HJM} is employed to solve
the differential equation for the action. In general, the most contributive
terms of this method are found in the leading order of the Planck constant $%
\hbar $ \cite{Mann1,Mann2,Jing,Majhi}. The emission and absorption
probabilities of the tunneling particles are given by \cite{Pad4}

\begin{equation}
\Gamma _{em}=\exp \left( -\frac{2Im S_{+}}{\hbar }\right) ,  \label{1n}
\end{equation}%
\begin{equation}
\Gamma _{ab}=\exp \left( -\frac{2 Im S_{-}}{\hbar }\right) ,  \label{2n}
\end{equation}

such that their ratio yields the tunneling probability as follows

\begin{equation}
\Gamma =\frac{\Gamma _{em}}{\Gamma _{ab}}=e^{-\beta E},  \label{3n}
\end{equation}%
where $\beta $ denotes the inverse temperature and $E$ is the energy \cite%
{Vanz}. Since the problem is nothing but the finding of the ratio (3), we
can always follow a way which normalizes the absorption probability to 1
(\%100 engulfment) thereof the classical description of the black hole. At
the same time, this classical condition on $\Gamma _{ab}$ enables us to read
the quantum mechanical output: the emission probability. Thus, the tunneling
probability becomes identical to the emission probability: $\Gamma =\Gamma
_{em}$. Similar to this methodology, there exists other ways to obtain $%
\Gamma $ (see for instance \cite{Akh1,Akh2}).

A first important step toward the understanding of elementary and composite
particles in a relativistic context was made by Klein \cite{Klein} and
Gordon \cite{Gordon}. Their resultant equation, the so-called Klein-Gordon
equation, is one of the powerful equations that describes the dynamics of
the relativistic spin-0 particles (scalar bosons, pions, scalar mesons, and
also Higgs particles) in any geometry. However, in nature many elementary
particles have spin and Dirac \cite{DiracBook} introduced the correct form
of the equation for fermions with spin-1/2 (such as electrons,quarks, and
leptons). Then, Proca \cite{proca} derived an equation for the massive
spin-1 particles (vector particles: $\mathcal{W}$ and $\mathcal{Z}$ bosons
and gluons) which play crucial role in the fundamental interactions as being
the force carriers: the $\mathcal{W}$ and $\mathcal{Z}$ bosons for the weak
interaction, and the gluons for the strong interaction. The Proca equation
reduces to the Maxwell equation with the limit of the zero masses (photons).
In the standard model, experimental verification of the fundamental
particles has only been made for the particles having spin 1 and less.
However, theoretical physics prescribes that graviton (massless boson) which
is the mediator of the gravitational force must have spin-2.

Nowadays, one of the trend subjects in physics is the massive gravity theory 
\cite{massive,bt,bt1}. This is a theory of gravity that modifies general
relativity by endowing the graviton with mass. In the classical theory, this
means that gravitational waves obey a massive wave equation and hence travel
at speeds below the speed of light \cite{arun}. A theory of a massive spin-2
field propagating in a spacetime background was discovered long ago by Fierz
and Pauli \cite{Fierz,Fierzalt}: the so-called Fierz-Pauli equations (FPEs).
However, it was understood in the 1970s that the massive graviton theories
have some drawbacks. These theories mainly suffer from the inevitable
inclusion of ghost modes and the problem of the general relativity limit
while the graviton's mass vanishes. Although some solutions to those
problems exists in three dimensions \cite{mg1,mg11,mg2}, they could not be
solved in four dimensions until the work of \ D'Amico et al \cite{mg3}. On
the other hand, LIGO and Virgo collaborations' observations have recently
confirmed the existence of gravitational waves \cite{LIGO}. Although these
experiments cannot detect individual gravitons, however they might provide
information about certain properties of the graviton.

Very recently, FPEs have been reformulated by Koenigstein et al \cite{spin2}
which is going to be employed in our present study for the derivation of the
HR arising from quantum tunneling of the massive gravitons through the event
horizon of a generic static and spherically symmetric (3+1) dimensional
black hole. To the best of our knowledge, this problem has not been studied
before. In this respect, the present study aims to fill this gap in the
literature.

\section{Quantum tunneling of massive spin-2 particles}

In this section, we will focus on the quantum tunneling of the massive
spin-2 bosons from a generic static and spherically symmetric (3+1)
dimensional black hole. To this end, we consider the metric as 
\begin{equation}
ds^{2}=-f(r)dt^{2}+\frac{1}{f(r)}dr^{2}+r^{2}\big[d\theta ^{2}+\left( \sin
\left( \theta \right) \right) ^{2}d\phi ^{2}\big].  \label{4n}
\end{equation}%
The horizon of such a black hole (4) is obtained via the condition of $%
f(r_{h})=0$.

According to the recent study of Koenigstein et al. \cite{spin2} [see its
equation (44)], the FPEs are given by

\begin{equation}
\mathcal{G}_{\mu \nu }=\square T_{\mu \nu }-\nabla _{\mu }\nabla _{\nu }T+%
\frac{m^{2}}{\hbar ^{2}}T_{\mu \nu }-\frac{m^{2}}{\hbar ^{2}}g_{\mu \nu
}T\,=0,  \label{5n}
\end{equation}%
where $\nabla _{\mu }$ denotes the covariant derivative, $\square$ is the
Laplacian operator:

\begin{equation}
\square =\nabla _{\mu }\nabla ^{\mu }=\frac{1}{\sqrt{-g}}\partial _{\nu
}\left( \sqrt{-g}\partial ^{\nu }\right) ,  \label{6n}
\end{equation}

and $T_{\mu \nu }$ represents the spin-2 fields (symmetric rank-2 tensor)
which can be expressed as \cite{eee}

\begin{equation}
T_{\mu \nu }=\nabla _{\mu }\mathit{\Phi}_{\nu }+\nabla _{\nu }\mathit{\Phi}%
_{\mu }-\frac{1}{2}g_{\mu \nu }\nabla ^{\beta }\mathit{\Phi}_{\beta },
\label{7n}
\end{equation}

in which $\mathit{\Phi}_{\nu }=\left(\mathit{\Phi}_{0},\mathit{\Phi}_{1},%
\mathit{\Phi}_{2},\mathit{\Phi}_{3}\right) $ stands for the two-component
spinor field. To obtain approximate solutions to the FPEs (5), we use the
following the WKB ansatz for the spinor field 
\begin{equation}
\mathit{\Phi} _{\nu }=\left( c_{0},c_{1},c_{2},c_{3}\right) e^{\frac{i}{%
\hbar }S(t,r,\theta ,\phi )},  \label{11n}
\end{equation}%
where $\left( c_{0},c_{1},c_{2},c_{3}\right) $ represent arbitrary
constants. The action seen in Eq. (8) is given by \cite{kruglov1} 
\begin{equation}
S(t,r,\theta ,\phi )=S_{0}(t,r,\theta ,\phi )+\hbar S_{1}(t,r,\theta ,\phi
)+\hbar ^{2}S_{2}(t,r,\theta ,\phi )+....\text{ .}  \label{12n}
\end{equation}%
After substituting Eqs. (8) and (9) in Eq. (5), we obtain sixteen equations
which can be seen in the Appendix. However, due to the symmetry Eqs.
(A1-A10) reduce a tetrad equation set that allows us to read the elements
(to the lowest order in $\hbar $) of the coefficient matrix $\Lambda \left(
c_{0},c_{1},c_{2},c_{3}\right) ^{T}=0$, where the superscript $T$ denotes
the transposition, as follows.

\begin{multline}
\Lambda_{11}=i\left( f \left( r \right) \right) ^{3} \left( {\frac {\partial 
}{ \partial \phi}}{S_{0}} \left( t,r,\theta,\phi \right) \right) ^{2}{\ 
\frac {\partial }{\partial r}}{S_{0}} \left( t,r,\theta,\phi \right) +i{r}%
^{2} \left( \sin \left( \theta \right) \right) ^{2} \left( f \left( r
\right) \right) ^{4} \left( {\frac {\partial }{\partial r}} {S_{0}} \left(
t,r,\theta,\phi \right) \right) ^{3}+ \\
i \left( f\left( r \right) \right) ^{3} \left( \sin \left( \theta \right)
\right) ^{2} \left( {\frac {\partial }{\partial r}}{S_{0}} \left( t,
r,\theta,\phi \right) \right) \left( {\frac {\partial }{\partial \theta}}{%
S_{0}} \left( t,r,\theta,\phi \right) \right) ^{2} -i\left( {\frac {\partial 
}{\partial r}}{S_{0}} \left( t,r,\theta,\phi \right) \right) {r}^{2}{m}^{2}
\left( f \left( r \right) \right) ^{ 3} \left( \sin \left( \theta \right)
\right) ^{2},
\end{multline}

\begin{multline}
\Lambda _{12}=i{r}^{2}\left( \sin \left( \theta \right) \right) ^{2}\left( {%
\ \frac{\partial }{\partial t}}{S_{0}}\left( t,r,\theta ,\phi \right)
\right) ^{3}-if\left( r\right) \left( \sin \left( \theta \right) \right)
^{2}\left( {\frac{\partial }{\partial t}}{S_{0}}\left( t,r,\theta ,\phi
\right) \right) \left( {\frac{\partial }{\partial \theta }}{S_{0}}\left(
t,r,\theta ,\phi \right) \right) ^{2}+ \\
i\left( {\frac{\partial }{\partial t}}{S_{0}}\left( t,r,\theta ,\phi \right)
\right) {r}^{2}{m}^{2}f\left( r\right) \left( \sin \left( \theta \right)
\right) ^{2}-if\left( r\right) \left( {\ \frac{\partial }{\partial \phi }}{%
S_{0}}\left( t,r,\theta ,\phi \right) \right) ^{2}{\frac{\partial }{\partial
t}}{S_{0}}\left( t,r,\theta ,\phi \right) ,
\end{multline}%
\begin{equation}
\Lambda _{13}=\Lambda _{14}=i{r}^{2}\left( \sin \left( \theta \right)
\right) ^{2}\left( f\left( r\right) \right) ^{2}\left( {\frac{\partial }{%
\partial r}}{S_{0}}\left( t,r,\theta ,\phi \right) \right) \left( {\frac{%
\partial }{\partial t}}{S_{0}}\left( t,r,\theta ,\phi \right) \right) {\frac{%
\partial }{\partial \theta }}{S_{0}}\left( t,r,\theta ,\phi \right) ,
\end{equation}

\begin{multline}
\Lambda_{21}= i{r}^{4} \left( \sin \left( \theta \right) \right) ^{4} \left(
f \left( r \right) \right) ^{3} \left( {\frac {\partial }{\partial r}} {S_{0}%
} \left( t,r,\theta,\phi \right) \right) ^{2}{\frac {\partial }{\partial
\theta}}{S_{0}} \left( t,r,\theta,\phi \right) +i{r}^{2} \left( \sin \left(
\theta \right) \right) ^{2} \left( f \left( r \right) \right) ^{2} \left( {%
\frac {\partial }{\partial \phi}}{S_{0}} \left( t,r,\theta,\phi \right)
\right) ^{2}\times \\
{\frac {\partial }{\partial \theta}} {S_{0}} \left( t,r,\theta,\phi \right)
+i{r}^{2} \left( \sin \left( \theta \right) \right) ^{4} \left( f \left( r
\right) \right) ^{2} \left( {\frac {\partial }{\partial \theta}}{\ S_{0}}
\left( t,r,\theta,\phi \right) \right) ^{3}-i \left( {\frac { \partial }{%
\partial \theta}}{S_{0}} \left( t,r,\theta,\phi \right) \right) {r}^{4}
\left( \sin \left( \theta \right) \right) ^{4}{m}^{2 } \left( f \left( r
\right) \right) ^{2},
\end{multline}

\begin{equation}
\Lambda _{22}=\Lambda _{24}=i{r}^{4}\left( \sin \left( \theta \right)
\right) ^{4}f\left( r\right) \left( {\frac{\partial }{\partial r}}{S_{0}}%
\left( t,r,\theta ,\phi \right) \right) \left( {\frac{\partial }{\partial t}}%
{\ S_{0}}\left( t,r,\theta ,\phi \right) \right) {\frac{\partial }{\partial
\theta }}{S_{0}}\left( t,r,\theta ,\phi \right) ,
\end{equation}

\begin{multline}
\Lambda_{23}= i{r}^{6} \left( \sin \left( \theta \right) \right) ^{4} \left( 
{\ \frac {\partial }{\partial t}}{S_{0}} \left( t,r,\theta,\phi \right)
\right) ^{3}-i{r}^{4} \left( \sin \left( \theta \right) \right) ^{2} f
\left( r \right) \left( {\frac {\partial }{\partial \phi}}{S_{0}} \left(
t,r,\theta,\phi \right) \right) ^{2}{\frac {\partial }{ \partial t}}{S_{0}}
\left( t,r,\theta,\phi \right) - \\
i{r}^{6} \left( \sin \left( \theta \right) \right) ^{4} \left( f \left( r
\right) \right) ^{2} \left( {\frac {\partial }{\partial r}}{S_{0}} \left( t,
r,\theta,\phi \right) \right) ^{2}{\frac {\partial }{\partial t}}{\ S_{0}}
\left( t,r,\theta,\phi \right) +i \left( {\frac {\partial }{ \partial t}}{%
S_{0}} \left( t,r,\theta,\phi \right) \right) {r}^{6} \left( \sin \left(
\theta \right) \right) ^{4}{m}^{2}f \left( r \right),
\end{multline}

\begin{multline}
\Lambda_{31}= i{r}^{2}\sin \left( \theta \right) \left( f \left( r \right)
\right) ^{2} \left( {\frac {\partial }{\partial \phi}}{S_{0}} \left(
t,r,\theta,\phi \right) \right) ^{3}+i{r}^{4} \left( \sin \left( \theta
\right) \right) ^{3} \left( f \left( r \right) \right) ^{3} \left( {\frac {%
\partial }{\partial \phi}}{S_{0}} \left( t,r,\theta,\phi \right) \right)
\left( {\frac {\partial }{ \partial r}}{S_{0}} \left( t,r,\theta,\phi
\right) \right) ^{2}+ \\
i{r}^{2} \left( \sin \left( \theta \right) \right) ^{3} \left( f \left( r
\right) \right) ^{2} \left( {\frac {\partial }{\partial \phi}}{S_{0}} \left(
t,r,\theta,\phi \right) \right) \left( {\frac {\partial } {\partial \theta}}{%
S_{0}} \left( t,r,\theta,\phi \right) \right) ^{2 }-i \left( {\frac {%
\partial }{\partial \phi}}{S_{0}} \left( t,r, \theta,\phi \right) \right) {r}%
^{4} \left( \sin \left( \theta \right) \right) ^{3}{m}^{2} \left( f \left( r
\right) \right) ^{2},
\end{multline}

\begin{equation}
\Lambda _{32}=\Lambda _{33}=i{r}^{4}\left( \sin \left( \theta \right)
\right) ^{3}f\left( r\right) \left( {\frac{\partial }{\partial \phi }}{S_{0}}%
\left( t,r,\theta ,\phi \right) \right) \left( {\frac{\partial }{\partial r}}%
{\ S_{0}}\left( t,r,\theta ,\phi \right) \right) {\frac{\partial }{\partial t%
}}{S_{0}}\left( t,r,\theta ,\phi \right) ,
\end{equation}

\begin{multline}
\Lambda_{34}=i{r}^{6} \left( \sin \left( \theta \right) \right) ^{5} \left( {%
\ \frac {\partial }{\partial t}}{S_{0}} \left( t,r,\theta,\phi \right)
\right) ^{3}-i{r}^{6} \left( \sin \left( \theta \right) \right) ^{5} \left(
f \left( r \right) \right) ^{2} \left( {\frac {\partial }{ \partial r}}{S_{0}%
} \left( t,r,\theta,\phi \right) \right) ^{2}{\ \frac {\partial }{\partial t}%
}{S_{0}} \left( t,r,\theta,\phi \right) + \\
i \left( {\frac {\partial }{\partial t}}{S_{0}} \left( t,r,\theta, \phi
\right) \right) {r}^{6} \left( \sin \left( \theta \right) \right) ^{5}{m}%
^{2}f \left( r \right) -i{r}^{4} \left( \sin \left( \theta \right) \right)
^{5}f \left( r \right) \left( {\frac { \partial }{\partial t}}{S_{0}} \left(
t,r,\theta,\phi \right) \right) \left( {\frac {\partial }{\partial \theta}}{%
S_{0}} \left( t ,r,\theta,\phi \right) \right) ^{2},
\end{multline}

\begin{equation}
\Lambda _{41}=\Lambda _{42}=i{r}^{4}\left( \sin \left( \theta \right)
\right) ^{3}f\left( r\right) \left( {\frac{\partial }{\partial \phi }}{S_{0}}%
\left( t,r,\theta ,\phi \right) \right) \left( {\frac{\partial }{\partial t}}%
{\ S_{0}}\left( t,r,\theta ,\phi \right) \right) {\frac{\partial }{\partial
\theta }}{S_{0}}\left( t,r,\theta ,\phi \right) ,
\end{equation}

\begin{multline}
\Lambda_{43}=i{r}^{6} \left( \sin \left( \theta \right) \right) ^{3} \left( {%
\ \frac {\partial }{\partial \phi}}{S_{0}} \left( t,r,\theta,\phi \right)
\right) \left( {\frac {\partial }{\partial t}}{S_{0}} \left( t,r,\theta,\phi
\right) \right) ^{2}-i{r}^{4}f \left( r \right) \sin \left( \theta \right)
\left( {\frac {\partial }{ \partial \phi}}{S_{0}} \left( t,r,\theta,\phi
\right) \right) ^{3}+ \\
i \left( {\frac {\partial }{\partial \phi}}{S_{0}} \left( t,r,\theta, \phi
\right) \right) {r}^{6}f \left( r \right) {m}^{2} \left( \sin \left( \theta
\right) \right) ^{3}-i{r}^{6} \left( f \left( r \right) \right) ^{2} \left(
\sin \left( \theta \right) \right) ^{3} \left( {\frac {\partial }{\partial
\phi}}{S_{0}} \left( t,r,\theta, \phi \right) \right) \left( {\frac {%
\partial }{\partial r}}{S_{0}} \left( t,r,\theta,\phi \right) \right) ^{2},
\end{multline}

\begin{multline}
\Lambda _{44}=i\left( {\frac{\partial }{\partial \theta }}{S_{0}}\left(
t,r,\theta ,\phi \right) \right) {r}^{6}\left( \sin \left( \theta \right)
\right) ^{5}{m}^{2}f\left( r\right) +i{r}^{6}\left( \sin \left( \theta
\right) \right) ^{5}\left( {\frac{\partial }{\partial t}}{S_{0}}\left(
t,r,\theta ,\phi \right) \right) ^{2}{\ \frac{\partial }{\partial \theta }}{%
S_{0}}\left( t,r,\theta ,\phi \right) - \\
i{r}^{4}\left( \sin \left( \theta \right) \right) ^{5}f\left( r\right)
\left( {\frac{\partial }{\partial \theta }}{S_{0}}\left( t,r,\theta ,\phi
\right) \right) ^{3}-i{r}^{6}\left( \sin \left( \theta \right) \right)
^{5}\left( f\left( r\right) \right) ^{2}\left( {\frac{\partial }{\partial r}}%
{S_{0}}\left( t,r,\theta ,\phi \right) \right) ^{2}{\frac{\partial }{%
\partial \theta }}{S_{0}}\left( t,r,\theta ,\phi \right) .
\end{multline}%
Having non-trivial solutions is conditional on $\mbox{\textit{det}}\Lambda =0
$ \cite{kruglov1,ia2} . Using the following Hamilton-Jacobi solution for the
lowest order action: 
\begin{equation}
S_{0}(t,r,\theta ,\phi )=-Et+W(r)+J\phi +K(\theta )+\Omega ,  \label{13n}
\end{equation}%
where $E$ and $J$ are the energy and angular momentum of the spin-2
particle, respectively, and $\Omega $ is a complex constant \cite{kruglov1},
the equation of $\mbox{\textit{det}}\Lambda =0$ becomes 
\begin{multline}
\left( f\left( r\right) \left( \sin \left( \theta \right) \right) ^{2}\left( 
{\frac{d}{d\theta }}K\left( \theta \right) \right) ^{2}+{r}^{2}\left( \sin
\left( \theta \right) \right) ^{2}\left( {\frac{d}{dr}}W\left( r\right)
\right) ^{2}\left( f\left( r\right) \right) ^{2}-{r}^{2}\left( {E}%
^{2}+f\left( r\right) {m}^{2}\right) \left( \sin \left( \theta \right)
\right) ^{2}+f\left( r\right) {J}^{2}\right) ^{3}\times  \\
\left( -f\left( r\right) \left( \sin \left( \theta \right) \right)
^{2}\left( {\frac{d}{d\theta }}K\left( \theta \right) \right) ^{2}-{r}%
^{2}\left( \sin \left( \theta \right) \right) ^{2}\left( {\frac{d}{dr}}%
W\left( r\right) \right) ^{2}\left( f\left( r\right) \right) ^{2}+{r}%
^{2}\left( {E}^{2}+2\,f\left( r\right) {m}^{2}\right) \left( \sin \left(
\theta \right) \right) ^{2}f\left( r\right) {J}^{2}\right) \times  \\
{r}^{10}\left( \sin \left( \theta \right) \right) ^{6}{E}^{2}=0.
\end{multline}%
From above, we obtain the following integral solutions for ${W}(r)$:%
\begin{equation}
W_{\pm }\left( r\right) =\pm \int {\frac{\sqrt{\left( -{J}^{2}-\left( \sin
\left( \theta \right) \right) ^{2}\left( {\frac{d}{d\theta }}K\left( \theta
\right) \right) ^{2}+{r}^{2}{m}^{2}\left( \sin \left( \theta \right) \right)
^{2}\right) f\left( r\right) +{\ r}^{2}\left( \sin \left( \theta \right)
\right) ^{2}{E}^{2}}}{r\sin \left( \theta \right) f\left( r\right) }},
\end{equation}%
and 
\begin{equation}
W_{\pm }\left( r\right) =\pm \int {\frac{\sqrt{\left( -{J}^{2}-\left( \sin
\left( \theta \right) \right) ^{2}\left( {\frac{d}{d\theta }}K\left( \theta
\right) \right) ^{2}+2\,{r}^{2}{m}^{2}\left( \sin \left( \theta \right)
\right) ^{2}\right) f\left( r\right) +{r}^{2}\left( \sin \left( \theta
\right) \right) ^{2}{E}^{2}}}{r\sin \left( \theta \right) f\left( r\right) }}%
.
\end{equation}%
where $+$ $(-)$ corresponds to the outgoing (ingoing) spin-2 particle. Here,
one can ask what would happen if we take the account of the van
Dam-Veltman-Zakharov (vDVZ) discontinuity \cite{vanDam,Zakharov} (the limit
of vanishing mass of the spin-2 particle) in the above integral solutions?
Since we shall consider Eqs. (24) and (25) around the
horizon, one can easily see that the expressions [including the mass ($m$) terms] which are the factors of the metric funtion  
$f\left( r\right) $\ in the square roots are
eliminated at the horizon: $f(r_{h})=0$. Namely, the zero-mass limit is
smooth during the quantum tunneling phenomenon. The latter
result is in agreement with the conclusion of the previous studies \cite%
{Kogan,Higuchi,Grisa}: "the vDVZ discontinuity is only present for a flat
background metric".

On the other hand, each $W_{\pm }(r)$ would have a simple pole at the
horizon. This pole problem can be overcome by employing the complex path
integration method prescribed in \cite{Pad4}. To this end, we first expand
the metric function $f(r)$ around the horizon: 
\begin{equation}
f(r)\approx \left. {\frac{d}{dr}}f\left( r\right) \right\vert _{r=r_{h}}+%
\mathcal{O}(r-r_{h})^{2}.  \label{26}
\end{equation}

Then, we insert Eq. (26) into integral solutions (24) and (25), and in
sequel evaluate the integrals. Thus, it can be seen that Eqs. (24) and (25)
give the same answer:

\begin{equation}
W_{\pm }(r)\approx \frac{\pm \pi iE}{\left. {\frac{d}{dr}}f\left( r\right)
\right\vert _{r=r_{h}}}.  \label{27}
\end{equation}

Hence, we infer from Eq. (22) that imaginary parts of the action come from
both Eq. (27) and the imaginary part of the complex constant $\Omega $. So,
the emission and absorption probabilities of the spin-2 particles crossing
the event horizon each way can be expressed as \cite{Mann1} 
\begin{equation}
\Gamma _{em}=\exp \left( -\frac{2Im S_{+}}{\hbar }\right)=\exp \left[ -\frac{%
2}{\hbar }\left( \mbox{Im}W_{+}+\mbox{Im}\Omega \right) \right] ,  \label{28}
\end{equation}%
\begin{equation}
\Gamma _{ab}=\exp \left( -\frac{2Im S_{-}}{\hbar }\right)=\exp \left[ -\frac{%
2}{\hbar }\left( \mbox{Im}W_{-}+\mbox{Im}\Omega \right) \right] .  \label{29}
\end{equation}%
According to the classical definition of the black hole, everything is
swallowed by it. This condition can be fulfilled by simply normalizing the
absorption probability to unity: $\Gamma _{ab}=1$. Thus, we see that $%
\mbox{Im}\Omega =-\mbox{Im}W_{-}$. On the other hand, since $W_{-}=-W_{+}$,
the latter condition is nothing short of $\mbox{Im}\Omega =\mbox{Im}W_{+}$,
and it whence yields the tunneling rate: 
\begin{eqnarray}
\Gamma &=&\frac{\Gamma _{em}}{\Gamma _{ab}}=\Gamma _{em},  \notag \\
&=&\exp \left( -\frac{4}{\hbar }\mbox{Im}W_{+}\right) ,  \notag \\
&=&\exp \left( \frac{-4\pi E}{\left. {\frac{d}{dr}}f\left( r\right)
\right\vert _{r=r_{h}}}\right) .  \label{30}
\end{eqnarray}%
Comparing Eq. (30) with Eq. (3), we read the radiation temperature which is
associated with the event horizon of the (3+1) dimensional BH as follows%
\begin{equation}
T=\frac{\hbar }{4\pi }\left. {\frac{d}{dr}}f\left( r\right) \right\vert
_{r=r_{h}}.  \label{31}
\end{equation}%
The above temperature is indeed the original Hawking temperature of the
static and symmetric black hole which is verified many times by the quantum
tunnelings of the other particles having spin rather than two \cite%
{Mann1,Mann2,kruglov2,Jing}. Namely, the radiation temperature obtained in
Eq. (31) is the reproduction of the standard Hawking temperature of the
metric (4).

\section{Conclusion}

In this study, we have studied the quantum tunneling of spin-2 massive
particles (bosons) from a generic (3+1) dimensional static and symmetric
black hole. By applying the WKB approximation (to the leading order in $%
\hbar $) and using the proper Hamilton-Jacobi ansatz in the FPEs [see Eqs.
(A1-A10)], we have managed to obtain the quantum tunneling rate (30) of the
massive spin-2 particles emitted from that generic black hole. The
corresponding black-body (from the Boltzmann formula) temperature (31) of
the radiation is matched with the well-known Hawking temperature of the
metric (4) which is recurrently verified by other particles' quantum
tunnelings. Meanwhile, during the quantum tunneling computations, it has been seen that the vDVZ discontinuity does not play role on the obtained tunneling rate (30).

It is worth noting that one can also compute the non-thermal radiation to
the radiation temperature by taking into account higher orders in $\hbar $.
However, such a study which is in our agenda requires more advanced
computations comparing with the present study, and its results are expected
to be remarkable from the aspect of information loss paradox \cite{Wil3}.
Moreover, the generalization of the present result to the other spacetimes
like the higher dimensional black holes, rotating black holes, dynamic black
holes, black strings, and wormholes may reveal more information rather than
this work. Those will also be our next problems in the near future.

\section*{Acknowledgments}

The authors are grateful to the editor and anonymous referee for their
valuable comments and suggestions to improve the paper.

\section{Appendix}

The sixteen FPEs (5) [due to the symmetry ($\mathcal{G}_{\alpha\beta}=%
\mathcal{G}_{\beta\alpha}$)] are as follows.

\begin{align*}
\mathcal{G}_{tt}& = & & \Bigg[\left( {\frac{\partial ^{3}}{\partial {t}^{3}}}%
\mathit{\Phi }_{0}\right) {r}^{2}\left( \sin \left( \theta \right) \right)
^{2}f\left( r\right) {\hbar }^{2}+2\,{\hbar }^{2}{r}^{2}\left( \sin \left(
\theta \right) \right) ^{2}\left( {\frac{\partial ^{2}}{\partial {t}^{2}}}%
\mathit{\Phi }_{1}\right) {\frac{d}{dr}}f\left( r\right) -{\hbar }^{2}{r}%
^{2}\left( \sin \left( \theta \right) \right) ^{2}\left( {\frac{d}{dr}}%
f\left( r\right) \right) ^{2} \\
& & & \times \left( {\frac{\partial }{\partial t}}\mathit{\Phi }_{0}\right)
f\left( r\right) -4\,{\hbar }^{2}{r}^{2}\left( \sin \left( \theta \right)
\right) ^{2}\left( {\frac{d}{dr}}f\left( r\right) \right) \left( f\left(
r\right) \right) ^{2}{\frac{\partial ^{2}}{\partial t\partial r}}\mathit{%
\Phi }_{0}+{\hbar }^{2}{r}^{2}\left( \sin \left( \theta \right) \right) ^{2}%
\mathit{\Phi }_{1}\left( {\frac{d}{dr}}f\left( r\right) \right) ^{3} \\
& & & -{\hbar }^{2}{r}^{2}\left( \sin \left( \theta \right) \right)
^{2}\left( {\frac{d}{dr}}f\left( r\right) \right) ^{2}\left( {\frac{\partial 
}{\partial r}}\mathit{\Phi }_{1}\right) f\left( r\right) -2{\hbar }%
^{2}\left( f\left( r\right) \right) ^{3}{r}^{2}\left( \sin \left( \theta
\right) \right) ^{2}{\frac{\partial ^{3}}{\partial t\partial {r}^{2}}}%
\mathit{\Phi }_{0}-2\,{\hbar }^{2}\left( f\left( r\right) \right) ^{2} \\
& & & \times \left( \sin \left( \theta \right) \right) ^{2}{\frac{\partial
^{3}}{\partial {\theta }^{2}\partial t}}\mathit{\Phi }_{0}-4{\hbar }%
^{2}\left( f\left( r\right) \right) ^{3}\left( \sin \left( \theta \right)
\right) ^{2}r{\frac{\partial ^{2}}{\partial t\partial r}}\mathit{\Phi }_{0}-2%
{\hbar }^{2}\left( f\left( r\right) \right) ^{2}{\frac{\partial ^{3}}{%
\partial t\partial {\phi }^{2}}}\mathit{\Phi }_{0}-2{\hbar }^{2}\left(
f\left( r\right) \right) ^{2} \\
& & & \times \sin \left( \theta \right) \cos \left( \theta \right) {\frac{%
\partial ^{2}}{\partial \theta \partial t}}\mathit{\Phi }_{0}-\left( {\frac{%
\partial ^{3}}{\partial {t}^{2}\partial r}}\mathit{\Phi }_{1}\right) {r}%
^{2}\left( \sin \left( \theta \right) \right) ^{2}f\left( r\right) {\hbar }%
^{2}-\left( {\frac{\partial ^{3}}{\partial \theta \partial {t}^{2}}}\mathit{%
\Phi }_{2}\right) {r}^{2}\left( \sin \left( \theta \right) \right)
^{2}f\left( r\right) {\hbar }^{2} \\
& & & -\left( {\frac{\partial ^{3}}{\partial {t}^{2}\partial \phi }}\mathit{%
\Phi }_{3}\right) {r}^{2}\left( \sin \left( \theta \right) \right)
^{2}f\left( r\right) {\hbar }^{2}-{m}^{2}\left( f\left( r\right) \right)
^{2}\left( {\frac{\partial }{\partial t}}\mathit{\Phi }_{0}\right) {r}%
^{2}\left( \sin \left( \theta \right) \right) ^{2}+{m}^{2}\left( f\left(
r\right) \right) ^{2}{r}^{2}\left( \sin \left( \theta \right) \right) ^{2} \\
& & & \times {\frac{\partial }{\partial r}}\mathit{\Phi }_{1}+{m}^{2}\left(
f\left( r\right) \right) ^{2}{r}^{2}\left( \sin \left( \theta \right)
\right) ^{2}{\frac{\partial }{\partial \theta }}\mathit{\Phi }_{2}+{m}%
^{2}\left( f\left( r\right) \right) ^{2}{r}^{2}\left( \sin \left( \theta
\right) \right) ^{2}{\frac{\partial }{\partial \phi }}\mathit{\Phi }_{3}%
\Bigg]\bigg/\bigg[{r}^{2}\left( \sin \left( \theta \right) \right)
^{2}f\left( r\right) {\hbar }^{2}\bigg]  \tag{A1}
\end{align*}

\begin{align*}
\mathcal{G}_{\theta t}=\mathcal{G}_{t \theta} &= & \Bigg[-2\,{\hbar }%
^{2}\left( \sin \left( \theta \right) \right) ^{3}{r}^{2}\left( {\frac{%
\partial }{\partial t}}\mathit{\Phi }_{2}\right) \left( f\left( r\right)
\right) ^{2}-2\,{\hbar }^{2}\left( \cos \left( \theta \right) \right)
^{2}\left( f\left( r\right) \right) ^{2}\left( {\frac{\partial }{\partial
\theta }}\mathit{\Phi }_{0}\right) \sin \left( \theta \right) -2\,{m}^{2}{r}%
^{4}  \notag \\
& & \times \left( \sin \left( \theta \right) \right) ^{3}f\left( r\right) {%
\frac{\partial }{\partial t}}\mathit{\Phi }_{2}+2\,{m}^{2}{r}^{2}\left( \sin
\left( \theta \right) \right) ^{3}\left( f\left( r\right) \right) ^{2}{\frac{%
\partial }{\partial \theta }}\mathit{\Phi }_{0}-2\,{\hbar }^{2}\left( \sin
\left( \theta \right) \right) ^{3}\left( f\left( r\right) \right) ^{3}{\frac{%
\partial }{\partial \theta }}\mathit{\Phi }_{0}  \notag \\
& & -4\,{\hbar }^{2}\cos \left( \theta \right) \left( f\left( r\right)
\right) ^{2}{\frac{\partial ^{2}}{\partial {\phi }^{2}}}\mathit{\Phi }%
_{0}+2\,{\hbar }^{2}\left( f\left( r\right) \right) ^{2}\left( {\frac{%
\partial ^{3}}{\partial \theta \partial {\phi }^{2}}}\mathit{\Phi }%
_{0}\right) \sin \left( \theta \right) +2\,{\hbar }^{2}\left( \sin \left(
\theta \right) \right) ^{3}  \notag \\
& & \left( f\left( r\right) \right) ^{2}{\frac{\partial ^{3}}{\partial {%
\theta }^{3}}}\mathit{\Phi }_{0}+2\,{\hbar }^{2}{r}^{4}\left( \sin \left(
\theta \right) \right) ^{3}{\frac{\partial ^{3}}{\partial {t}^{3}}}\mathit{%
\Phi }_{2}+{\hbar }^{2}\left( \sin \left( \theta \right) \right) ^{3}\left(
f\left( r\right) \right) ^{2}{r}^{2}\left( {\frac{d^{2}}{d{r}^{2}}}f\left(
r\right) \right) {\frac{\partial }{\partial \theta }}\mathit{\Phi }_{0} 
\notag \\
& & +{\hbar }^{2}\left( \sin \left( \theta \right) \right) ^{3}f\left(
r\right) {r}^{4}\left( {\frac{d^{2}}{d{r}^{2}}}f\left( r\right) \right) {%
\frac{\partial }{\partial t}}\mathit{\Phi }_{2}+4\,{\hbar }^{2}{r}^{2}\left(
\sin \left( \theta \right) \right) ^{3}\left( {\frac{d}{dr}}f\left( r\right)
\right) \left( f\left( r\right) \right) ^{2}{\frac{\partial ^{2}}{\partial
\theta \partial r}}\mathit{\Phi }_{0}  \notag \\
& & -2\,{\hbar }^{2}f\left( r\right) {r}^{2}\left( {\frac{\partial ^{2}}{%
\partial \theta \partial t}}\mathit{\Phi }_{2}\right) \left( \sin \left(
\theta \right) \right) ^{2}\cos \left( \theta \right) +4\,{\hbar }%
^{2}f\left( r\right) {r}^{2}\left( \sin \left( \theta \right) \right)
^{2}\left( {\frac{\partial ^{2}}{\partial t\partial \phi }}\mathit{\Phi }%
_{3}\right) \cos \left( \theta \right)  \notag \\
& & -2\,{\hbar }^{2}{r}^{4}\left( \sin \left( \theta \right) \right)
^{3}\left( {\frac{d}{dr}}f\left( r\right) \right) \left( {\frac{\partial ^{2}%
}{\partial t\partial r}}\mathit{\Phi }_{2}\right) f\left( r\right) -4\,{%
\hbar }^{2}\left( \sin \left( \theta \right) \right) ^{3}f\left( r\right) r{%
\frac{\partial ^{2}}{\partial \theta \partial t}}\mathit{\Phi }_{1}-2\,{%
\hbar }^{2}\left( \sin \left( \theta \right) \right) ^{3}  \notag \\
& & \times \left( f\left( r\right) \right) ^{2}{r}^{4}{\frac{\partial ^{3}}{%
\partial t\partial {r}^{2}}}\mathit{\Phi }_{2}+2{\hbar }^{2}\left( \sin
\left( \theta \right) \right) ^{3}\left( f\left( r\right) \right) ^{3}{r}^{2}%
{\frac{\partial ^{3}}{\partial \theta \partial {r}^{2}}}\mathit{\Phi }%
_{0}+4\,{\hbar }^{2}\left( \sin \left( \theta \right) \right) ^{3}\left(
f\left( r\right) \right) ^{3}\left( {\frac{\partial ^{2}}{\partial \theta
\partial r}}\mathit{\Phi }_{0}\right) r  \notag \\
& & -2\,{h}^{2}{r}^{2}\left( \sin \left( \theta \right) \right) ^{3}\left( {%
\frac{\partial ^{2}}{\partial \theta \partial t}}\mathit{\Phi }_{1}\right) {%
\frac{d}{dr}}f\left( r\right) -8\,{h}^{2}\left( \sin \left( \theta \right)
\right) ^{3}\left( f\left( r\right) \right) ^{2}{r}^{3}{\frac{\partial ^{2}}{%
\partial t\partial r}}\mathit{\Phi }_{2}+2\,\left( {\frac{\partial ^{3}}{%
\partial \theta \partial t\partial \phi }}\mathit{\Phi }_{3}\right)  \notag
\\
& & \times {r}^{2}\left( \sin \left( \theta \right) \right) ^{3}f\left(
r\right) {h}^{2}+2\,\left( {\frac{\partial ^{3}}{\partial \theta \partial
t\partial r}}\mathit{\Phi }_{1}\right) {r}^{2}\left( \sin \left( \theta
\right) \right) ^{3}f\left( r\right) {h}^{2}-2\,{h}^{2}f\left( r\right) {r}%
^{2}\left( {\frac{\partial ^{3}}{\partial t\partial {\phi }^{2}}}\mathit{%
\Phi }_{2}\right) \sin \left( \theta \right)  \notag \\
& & +2\,{h}^{2}\left( \sin \left( \theta \right) \right) ^{2}\cos \left(
\theta \right) \left( f\left( r\right) \right) ^{2}{\frac{\partial ^{2}}{%
\partial {\theta }^{2}}}\mathit{\Phi }_{0}-2\,{h}^{2}{r}^{3}\left( \sin
\left( \theta \right) \right) ^{3}\left( {\frac{d}{dr}}f\left( r\right)
\right) \left( {\frac{\partial }{\partial t}}\mathit{\Phi }_{2}\right)
f\left( r\right)  \notag \\
& & +2\,{h}^{2}r\left( \sin \left( \theta \right) \right) ^{3}\left( {\frac{d%
}{dr}}f\left( r\right) \right) \left( f\left( r\right) \right) ^{2}{\frac{%
\partial }{\partial \theta }}\mathit{\Phi }_{0}+2\,{h}^{2}f\left(r\right)
\left( \cos \left( \theta \right) \right) ^{2} {r}^{2}\left( {\frac{\partial 
}{\partial t}}\mathit{\Phi }_{2}\right) \sin \left( \theta \right) \Bigg] 
\notag \\
& & \bigg/\bigg[-2{r}^{2}\left( \sin \left( \theta \right) \right)
^{3}f\left( r\right) {\hbar }^{2})\bigg]  \tag{A2}
\end{align*}

\begin{align*}
\mathcal{G}_{\phi t}=\mathcal{G}_{t\phi } &= & \Bigg[2\,{\hbar }^{2}\left(
f\left( r\right) \right) ^{2}\left( {\frac{\partial }{\partial \phi }}%
\mathit{\Phi }_{0}\right) \left( \sin \left( \theta \right) \right) ^{2}-2\,{%
\hbar }^{2}\left( \sin \left( \theta \right) \right) ^{2}\left( f\left(
r\right) \right) ^{3}{\frac{\partial }{\partial \phi }}\mathit{\Phi }_{0}+2\,%
{\hbar }^{2}\left( \sin \left( \theta \right) \right) ^{2}\left( f\left(
r\right) \right) ^{2} \\
& & \times {\frac{\partial ^{3}}{\partial {\theta }^{2}\partial \phi }}%
\mathit{\Phi }_{0}+2\,{\hbar }^{2}{r}^{4}\left( \sin \left( \theta \right)
\right) ^{4}{\frac{\partial ^{3}}{\partial {t}^{3}}}\mathit{\Phi }_{3}+2\,{%
\hbar }^{2}{r}^{2}\left( \sin \left( \theta \right) \right) ^{4}\left( {%
\frac{\partial }{\partial t}}\mathit{\Phi }_{3}\right) f\left( r\right) -2\,{%
\hbar }^{2}{r}^{2}\left( \sin \left( \theta \right) \right) ^{4} \\
& & \times \left( {\frac{\partial }{\partial t}}\mathit{\Phi }_{3}\right)
\left( f\left( r\right) \right) ^{2}-2\,{m}^{2}{r}^{4}\left( \sin \left(
\theta \right) \right) ^{4}f\left( r\right) {\frac{\partial }{\partial t}}%
\mathit{\Phi }_{3}+2\,{m}^{2}{r}^{2}\left( \sin \left( \theta \right)
\right) ^{2}\left( f\left( r\right) \right) ^{2}{\frac{\partial }{\partial
\phi }}\mathit{\Phi }_{0} \\
& & +2\,\left( {\frac{\partial ^{3}}{\partial t\partial r\partial \phi }}%
\mathit{\Phi }_{1}\right) {r}^{2}\left( \sin \left( \theta \right) \right)
^{2}f\left( r\right) {\hbar }^{2}+4\,{\hbar }^{2}\left( \sin \left( \theta
\right) \right) ^{2}\left( f\left( r\right) \right) ^{3}\left( {\frac{%
\partial ^{2}}{\partial r\partial \phi }}\mathit{\Phi }_{0}\right) r+2\,{%
\hbar }^{2}\left( \sin \left( \theta \right) \right) ^{2} \\
& & \times \left( f\left( r\right) \right) ^{3}{r}^{2}{\frac{\partial ^{3}}{%
\partial {r}^{2}\partial \phi }}\mathit{\Phi }_{0}-2\,{\hbar }^{2}\left(
\sin \left( \theta \right) \right) ^{4}\left( f\left( r\right) \right) ^{2}{r%
}^{4}{\frac{\partial ^{3}}{\partial t\partial {r}^{2}}}\mathit{\Phi }_{3}-8\,%
{\hbar }^{2}\left( \sin \left( \theta \right) \right) ^{4}\left( f\left(
r\right) \right) ^{2}{r}^{3} \\
& & \times {\frac{\partial ^{2}}{\partial t\partial r}}\mathit{\Phi }_{3}-2\,%
{\hbar }^{2}{r}^{2}\left( \sin \left( \theta \right) \right) ^{2}\left( {%
\frac{\partial ^{2}}{\partial t\partial \phi }}\mathit{\Phi }_{1}\right) {%
\frac{d}{dr}}f\left( r\right) +2\,\left( {\frac{\partial ^{3}}{\partial
\theta \partial t\partial \phi }}\mathit{\Phi }_{2}\right) {r}^{2}\left(
\sin \left( \theta \right) \right) ^{2}f\left( r\right) {\hbar }^{2} \\
& & -2\,{\hbar }^{2}f\left( r\right) \left( \sin \left( \theta \right)
\right) ^{4}{r}^{2}{\frac{\partial ^{3}}{\partial {\theta }^{2}\partial t}}%
\mathit{\Phi }_{3}-4\,{\hbar }^{2}f\left( r\right) r\left( \sin \left(
\theta \right) \right) ^{2}{\frac{\partial ^{2}}{\partial t\partial \phi }}%
\mathit{\Phi }_{1}+2\,{\hbar }^{2}\cos \left( \theta \right) \left( f\left(
r\right) \right) ^{2} \\
& & \times \left( {\frac{\partial ^{2}}{\partial \theta \partial \phi }}%
\mathit{\Phi }_{0}\right) \sin \left( \theta \right) +2\,{\hbar }^{2}\left(
f\left( r\right) \right) ^{2}{\frac{\partial ^{3}}{\partial {\phi }^{3}}}%
\mathit{\Phi }_{0}-2\,{\hbar }^{2}{r}^{3}\left( \sin \left( \theta \right)
\right) ^{4}\left( {\frac{d}{dr}}f\left( r\right) \right) \left( {\frac{%
\partial }{\partial t}}\mathit{\Phi }_{3}\right) f\left( r\right) \\
& & +2\,{\hbar }^{2}r\left( \sin \left( \theta \right) \right) ^{2}\left( {%
\frac{d}{dr}}f\left( r\right) \right) \left( f\left( r\right) \right) ^{2}{%
\frac{\partial }{\partial \phi }}\mathit{\Phi }_{0}+{\hbar }^{2}\left( \sin
\left( \theta \right) \right) ^{4}f\left( r\right) {r}^{4}\left( {\frac{d^{2}%
}{d{r}^{2}}}f\left( r\right) \right) {\frac{\partial }{\partial t}}\mathit{%
\Phi }_{3} \\
& & +{\hbar }^{2}\left( \sin \left( \theta \right) \right) ^{2}\left(
f\left( r\right) \right) ^{2}{r}^{2}\left( {\frac{d^{2}}{d{r}^{2}}}f\left(
r\right) \right) {\frac{\partial }{\partial \phi }}\mathit{\Phi }_{0}-6\,{%
\hbar }^{2}f\left( r\right) \left( \sin \left( \theta \right) \right)
^{3}\cos \left( \theta \right) {r}^{2} \\
& & \times \frac{\partial ^{2}}{\partial \theta \partial t}\mathit{\Phi }%
_{3}+4\,{\hbar }^{2}{r}^{2}\left( \sin \left( \theta \right) \right)
^{2}\left( {\frac{d}{dr}}f\left( r\right) \right) \left( f\left( r\right)
\right) ^{2}{\frac{\partial ^{2}}{\partial r\partial \phi }}\mathit{\Phi }%
_{0}-4\,{\hbar }^{2}f\left( r\right) \sin \left( \theta \right) \cos \left(
\theta \right) {r}^{2} \\
& & \times {\frac{\partial ^{2}}{\partial t\partial \phi }}\mathit{\Phi }%
_{2}-2\,{\hbar }^{2}{r}^{4}\left( \sin \left( \theta \right) \right)
^{4}\left( {\frac{d}{dr}}f\left( r\right) \right) \left( {\frac{\partial ^{2}%
}{\partial t\partial r}}\mathit{\Phi }_{3}\right) f\left( r\right) \Bigg]%
\bigg/\bigg[-2{r}^{2}\left( \sin \left( \theta \right) \right) ^{2}f\left(
r\right) {\hbar }^{2}\bigg]  \tag{A3}
\end{align*}

\begin{align*}
\mathcal{G}_{r\theta }=\mathcal{G}_{\theta r}& = & & \Bigg[10\,{\hbar }%
^{2}\left( \sin \left( \theta \right) \right) ^{3}\left( f\left( r\right)
\right) ^{2}{\frac{\partial }{\partial \theta }}\mathit{\Phi }_{1}+2\,{\hbar 
}^{2}{r}^{2}\left( \sin \left( \theta \right) \right) ^{3}{\frac{\partial
^{3}}{\partial \theta \partial {t}^{2}}}\mathit{\Phi }_{1}+4\,{\hbar }%
^{2}f\left( r\right) \cos \left( \theta \right) \\
& & & \times {\frac{\partial ^{2}}{\partial {\phi }^{2}}}\mathit{\Phi }%
_{1}-2\,{\hbar }^{2}f\left( r\right) \sin \left( \theta \right) {\frac{%
\partial ^{3}}{\partial \theta \partial {\phi }^{2}}}\mathit{\Phi }_{1}-2\,{%
\hbar }^{2}\left( \sin \left( \theta \right) \right) ^{3}f\left( r\right) {%
\frac{\partial ^{3}}{\partial {\theta }^{3}}}\mathit{\Phi }_{1} \\
& & & +2\,\left( {\frac{\partial ^{3}}{\partial \theta \partial r\partial
\phi }}\mathit{\Phi }_{3}\right) \left( f\left( r\right) \right) ^{2}{r}%
^{2}\left( \sin \left( \theta \right) \right) ^{3}{\hbar }^{2}+4\,{\hbar }%
^{2}{r}^{3}\left( \sin \left( \theta \right) \right) ^{3}\left( {\frac{%
\partial ^{2}}{\partial {t}^{2}}}\mathit{\Phi }_{2}\right) f\left( r\right)
+2{\hbar }^{2}{r}^{4}\left( \sin \left( \theta \right) \right) ^{3} \\
& & & \times \left( {\frac{\partial ^{3}}{\partial {t}^{2}\partial r}}%
\mathit{\Phi }_{2}\right) f\left( r\right) -2{\hbar }^{2}{r}^{4}\left( \sin
\left( \theta \right) \right) ^{3}\left( {\frac{d}{dr}}f\left( r\right)
\right) {\frac{\partial ^{2}}{\partial {t}^{2}}}\mathit{\Phi }_{2}+20\,{%
\hbar }^{2}\left( \sin \left( \theta \right) \right) ^{3}\left( f\left(
r\right) \right) ^{3}r\mathit{\Phi }_{2} \\
& & & -2{\hbar }^{2}\left( \sin \left( \theta \right) \right) ^{3}\left(
f\left( r\right) \right) ^{3}{r}^{2}{\frac{\partial }{\partial r}}\mathit{%
\Phi }_{2}+2\,{\hbar }^{2}f\left( r\right) \sin \left( \theta \right) \left(
\cos \left( \theta \right) \right) ^{2}{\frac{\partial }{\partial \theta }}%
\mathit{\Phi }_{1}-4\,{m}^{2}{r}^{3}\left( \sin \left( \theta \right)
\right) ^{3}\left( f\left( r\right) \right) ^{2}\mathit{\Phi }_{2} \\
& & & -2\,{m}^{2}{r}^{4}\left( \sin \left( \theta \right) \right) ^{3}\left(
f\left( r\right) \right) ^{2}{\frac{\partial }{\partial r}}\mathit{\Phi }%
_{2}-2\,{m}^{2}{r}^{2}\left( \sin \left( \theta \right) \right) ^{3}f\left(
r\right) {\frac{\partial }{\partial \theta }}\mathit{\Phi }_{1}-2\,{\hbar }%
^{2}\sin \left( \theta \right) {r}^{2} \\
& & & \times \left( {\frac{\partial ^{3}}{\partial r\partial {\phi }^{2}}}%
\mathit{\Phi }_{2}\right) \left( f\left( r\right) \right) ^{2}+4\,{\hbar }%
^{2}\left( \sin \left( \theta \right) \right) ^{3}\left( f\left( r\right)
\right) ^{2}r{\frac{\partial ^{2}}{\partial {\theta }^{2}}}\mathit{\Phi }%
_{2}+4\,{\hbar }^{2}\left( f\left( r\right) \right) ^{2}\left( \sin \left(
\theta \right) \right) ^{3}r{\frac{\partial ^{2}}{\partial \theta \partial
\phi }}\mathit{\Phi }_{2} \\
& & & -2\,{\hbar }^{2}f\left( r\right) \left( \sin \left( \theta \right)
\right) ^{2}\cos \left( \theta \right) {\frac{\partial ^{2}}{\partial {%
\theta }^{2}}}\mathit{\Phi }_{1}-8\,{\hbar }^{2}\left( \sin \left( \theta
\right) \right) ^{3}\left( f\left( r\right) \right) ^{2}r{\frac{\partial ^{2}%
}{\partial \theta \partial r}}\mathit{\Phi }_{1}-2\,{\hbar }^{2}\left( \sin
\left( \theta \right) \right) ^{3}\left( f\left( r\right) \right) ^{3} \\
& & & \times {r}^{4}{\frac{\partial ^{3}}{\partial {r}^{3}}}\mathit{\Phi }%
_{2}-12\,{\hbar }^{2}\left( \sin \left( \theta \right) \right) ^{3}\left(
f\left( r\right) \right) ^{3}{r}^{3}{\frac{\partial ^{2}}{\partial {r}^{2}}}%
\mathit{\Phi }_{2}+2\,\left( {\frac{\partial ^{3}}{\partial \theta \partial
t\partial r}}\mathit{\Phi }_{0}\right) \left( f\left( r\right) \right) ^{2}{r%
}^{2}\left( \sin \left( \theta \right) \right) ^{3}{\hbar }^{2} \\
& & & -14\,{\hbar }^{2}{r}^{3}\left( \sin \left( \theta \right) \right)
^{3}\left( {\frac{d}{dr}}f\left( r\right) \right) \left( f\left( r\right)
\right) ^{2}{\frac{\partial }{\partial r}}\mathit{\Phi }_{2}+10\,{\hbar }%
^{2}r\left( \sin \left( \theta \right) \right) ^{3}\left( {\frac{d}{dr}}%
f\left( r\right) \right) f\left( r\right) {\frac{\partial }{\partial \theta }%
}\mathit{\Phi }_{1} \\
& & & -4{\hbar }^{2}\left( \sin \left( \theta \right) \right) ^{3}\left(
f\left( r\right) \right) ^{2}{r}^{2}\mathit{\Phi }_{2}{\frac{d}{dr}}f\left(
r\right) +4\,{\hbar }^{2}\sin \left( \theta \right) \left( \cos \left(
\theta \right) \right) ^{2}r\mathit{\Phi }_{2}\left( f\left( r\right)
\right) ^{2}+4\,{\hbar }^{2}\left( \sin \left( \theta \right) \right) ^{2} \\
& & & \times \cos \left( \theta \right) r\left( {\frac{\partial }{\partial
\theta }}\mathit{\Phi }_{2}\right) \left( f\left( r\right) \right) ^{2}+2\,{%
\hbar }^{2}\sin \left( \theta \right) \left( \cos \left( \theta \right)
\right) ^{2}{r}^{2}\left( {\frac{\partial }{\partial r}}\mathit{\Phi }%
_{2}\right) \left( f\left( r\right) \right) ^{2}+8{\hbar }^{2}r\left( \sin
\left( \theta \right) \right) ^{2} \\
& & & \times \left( {\frac{\partial }{\partial \phi }}\mathit{\Phi }%
_{3}\right) \left( f\left( r\right) \right) ^{2}\cos \left( \theta \right) -{%
\hbar }^{2}\left( \sin \left( \theta \right) \right) ^{3}\left( f\left(
r\right) \right) ^{2}{r}^{4}\left( {\frac{\partial }{\partial r}}\mathit{%
\Phi }_{2}\right) {\frac{d^{2}}{d{r}^{2}}}f\left( r\right) \\
& & & +{\hbar }^{2}\left( \sin \left( \theta \right) \right) ^{3}f\left(
r\right) {r}^{2}\left( {\frac{d^{2}}{d{r}^{2}}}f\left( r\right) \right) {%
\frac{\partial }{\partial \theta }}\mathit{\Phi }_{1}-4\,{\hbar }^{2}{r}%
^{4}\left( \sin \left( \theta \right) \right) ^{3}\left( {\frac{d}{dr}}%
f\left( r\right) \right) \left( f\left( r\right) \right) ^{2}{\frac{\partial
^{2}}{\partial {r}^{2}}}\mathit{\Phi }_{2} \\
& & & -2\,{\hbar }^{2}\left( \sin \left( \theta \right) \right) ^{3}\left(
f\left( r\right) \right) ^{2}{r}^{3}\mathit{\Phi }_{2}{\frac{d^{2}}{d{r}^{2}}%
}f\left( r\right) +4\,{\hbar }^{2}{r}^{2}\left( \sin \left( \theta \right)
\right) ^{2}\left( {\frac{\partial ^{2}}{\partial r\partial \phi }}\mathit{%
\Phi }_{3}\right) \left( f\left( r\right) \right) ^{2}\cos \left( \theta
\right) \\
& & & -2\,{\hbar }^{2}\left( \sin \left( \theta \right) \right) ^{2}\cos
\left( \theta \right) {r}^{2}\left( {\frac{\partial ^{2}}{\partial \theta
\partial r}}\mathit{\Phi }_{2}\right) \left( f\left( r\right) \right)
^{2}+2\,{\hbar }^{2}{r}^{2}\left( \sin \left( \theta \right) \right)
^{3}\left( {\frac{d}{dr}}f\left( r\right) \right) \left( {\frac{\partial ^{2}%
}{\partial \theta \partial t}}\mathit{\Phi }_{0}\right) \\
& & & \times f\left( r\right) \Bigg]\bigg/\bigg[-2{\hbar }^{2}\left( \sin
\left( \theta \right) \right) ^{3}\left( f\left( r\right) \right) ^{2}{r}^{2}%
\bigg]  \tag{A4}
\end{align*}

\begin{align*}
\mathcal{G}_{\theta \theta } &= & \Bigg[2\,{\hbar }^{2}{r}^{3}\left( \sin
\left( \theta \right) \right) ^{3}{\frac{\partial ^{3}}{\partial \theta
\partial {t}^{2}}}\mathit{\Phi }_{2}-2\,{\hbar }^{2}{r}^{3}\left( \sin
\left( \theta \right) \right) ^{3}\left( {\frac{\partial ^{2}}{\partial
\theta \partial r}}\mathit{\Phi }_{2}\right) f\left( r\right) {\frac{d}{dr}}%
f\left( r\right) -2\,{\hbar }^{2}\left( f\left( r\right) \right) ^{2}{r}%
^{3}\left( \sin \left( \theta \right) \right) ^{3}  \notag \\
& & {\frac{\partial ^{3}}{\partial \theta \partial {r}^{2}}}\mathit{\Phi }%
_{2}-\left( {\frac{\partial ^{3}}{\partial {\theta }^{3}}}\mathit{\Phi }%
_{2}\right) r\left( \sin \left( \theta \right) \right) ^{3}f\left( r\right) {%
\hbar }^{2}-4\,{\hbar }^{2}\left( \sin \left( \theta \right) \right)
^{3}\left( f\left( r\right) \right) ^{2}r{\frac{\partial }{\partial \theta }}%
\mathit{\Phi }_{2}  \notag \\
& & -8\,{\hbar }^{2}\left( \sin \left( \theta \right) \right) ^{3}\left(
f\left( r\right) \right) ^{2}{r}^{2}{\frac{\partial ^{2}}{\partial \theta
\partial r}}\mathit{\Phi }_{2}-4\,{\hbar }^{2}\left( \sin \left( \theta
\right) \right) ^{3}f\left( r\right) {\frac{\partial ^{2}}{\partial {\theta }%
^{2}}}\mathit{\Phi }_{1}+4{\hbar }^{2}\left( \sin \left( \theta \right)
\right) ^{3}f\left( r\right) r\mathit{\Phi }_{1}  \notag \\
& & \times {\frac{d}{dr}}f\left( r\right) -4\,{\hbar }^{2}\left( \sin \left(
\theta \right) \right) ^{3}\left( f\left( r\right) \right) ^{2}r{\frac{%
\partial }{\partial r}}\mathit{\Phi }_{1}-2\,{\hbar }^{2}f\left( r\right)
r\left( {\frac{\partial ^{3}}{\partial \theta \partial {\phi }^{2}}}\mathit{%
\Phi }_{2}\right) \sin \left( \theta \right) +4\,{\hbar }^{2}f\left(
r\right) r\sin \left( \theta \right)  \notag \\
& & \times \left( {\frac{\partial }{\partial \phi }}\mathit{\Phi }%
_{3}\right) \left( \cos \left( \theta \right) \right) ^{2}+4\,{\hbar }%
^{2}f\left( r\right) r\cos \left( \theta \right) \left( \sin \left( \theta
\right) \right) ^{2}{\frac{\partial ^{2}}{\partial \theta \partial \phi }}%
\mathit{\Phi }_{3}+4{\hbar }^{2}f\left( r\right) r\cos \left( \theta \right)
\notag \\
& & \times {\frac{\partial ^{2}}{\partial {\phi }^{2}}}\mathit{\Phi }_{2}-2\,%
{\hbar }^{2}f\left( r\right) r\left( {\frac{\partial ^{2}}{\partial {\theta }%
^{2}}}\mathit{\Phi }_{2}\right) \left( \sin \left( \theta \right) \right)
^{2}\cos \left( \theta \right) +4\,{\hbar }^{2}f\left( r\right) r\left( {%
\frac{\partial }{\partial \theta }}\mathit{\Phi }_{2}\right) \left( \cos
\left( \theta \right) \right) ^{2}\sin \left( \theta \right)  \notag \\
& & +\left( {\frac{\partial ^{3}}{\partial {\theta }^{2}\partial t}}\mathit{%
\Phi }_{0}\right) r\left( \sin \left( \theta \right) \right) ^{3}f\left(
r\right) {\hbar }^{2}+\left( {\frac{\partial ^{3}}{\partial {\theta }%
^{2}\partial r}}\mathit{\Phi }_{1}\right) r\left( \sin \left( \theta \right)
\right) ^{3}f\left( r\right) {\hbar }^{2}+\left( {\frac{\partial ^{3}}{%
\partial {\theta }^{2}\partial \phi }}\mathit{\Phi }_{3}\right)  \notag \\
& & \times r\left( \sin \left( \theta \right) \right) ^{3}f\left( r\right) {%
\hbar }^{2}-{m}^{2}{r}^{3}\left( {\frac{\partial }{\partial \theta }}\mathit{%
\Phi }_{2}\right) \left( \sin \left( \theta \right) \right) ^{3}f\left(
r\right) +{m}^{2}{r}^{3}\left( \sin \left( \theta \right) \right)
^{3}f\left( r\right) {\frac{\partial }{\partial t}}\mathit{\Phi }_{0}+{m}^{2}%
{r}^{3}\left( \sin \left( \theta \right) \right) ^{3}  \notag \\
& & f\left( r\right) {\frac{\partial }{\partial r}}\mathit{\Phi }_{1}+{m}^{2}%
{r}^{3}\left( \sin \left( \theta \right) \right) ^{3}f\left( r\right) {\frac{%
\partial }{\partial \phi }}\mathit{\Phi }_{3}\Bigg]\bigg/\bigg[-{\hbar }%
^{2}\left( \sin \left( \theta \right) \right) ^{3}f\left( r\right) r\bigg] 
\tag{A5}
\end{align*}

\begin{align*}
\mathcal{G}_{\phi \phi } &= & \Bigg[4\,{\hbar }^{2}\left( \sin \left( \theta
\right) \right) ^{3}f\left( r\right) r\mathit{\Phi }_{1}{\frac{d}{dr}}%
f\left( r\right) -4\,{\hbar }^{2}f\left( r\right) r\sin \left( \theta
\right) \left( {\frac{\partial }{\partial \phi }}\mathit{\Phi }_{3}\right)
\left( \cos \left( \theta \right) \right) ^{2}-6\,{\hbar }^{2}f\left(
r\right) r\cos \left( \theta \right) \left( \sin \left( \theta \right)
\right) ^{2}  \notag \\
& & \times {\frac{\partial ^{2}}{\partial \theta \partial \phi }}\mathit{%
\Phi }_{3}-4\,{\hbar }^{2}f\left( r\right) r\left( {\frac{\partial }{%
\partial \theta }}\mathit{\Phi }_{2}\right) \left( \cos \left( \theta
\right) \right) ^{2}\sin \left( \theta \right) -8\,{\hbar }^{2}r\left( \sin
\left( \theta \right) \right) ^{2}\mathit{\Phi }_{2}\left( f\left( r\right)
\right) ^{2}\cos \left( \theta \right)  \notag \\
& & -4\,{\hbar }^{2}{r}^{2}\left( \sin \left( \theta \right) \right)
^{2}\left( {\frac{\partial }{\partial r}}\mathit{\Phi }_{2}\right)\left(
f\left( r\right) \right) ^{2}\cos \left( \theta \right) -2\,\left( {\frac{%
\partial ^{3}}{\partial {\theta }^{2}\partial \phi }}\mathit{\Phi }%
_{3}\right) r\left( \sin \left( \theta \right) \right) ^{3}f\left( r\right) {%
\hbar }^{2}  \notag \\
& & -4\,{\hbar }^{2}\left( \sin \left( \theta \right) \right) ^{3}\left(
f\left( r\right) \right) ^{2}r{\frac{\partial }{\partial r}}\mathit{\Phi }%
_{1}+{\hbar }^{2}f\left( r\right) r\left( {\frac{\partial ^{3}}{\partial
\theta \partial {\phi }^{2}}}\mathit{\Phi }_{2}\right) \sin \left( \theta
\right)  \notag \\
& & -4{\hbar }^{2}f\left( r\right) r\cos \left( \theta \right) {\frac{%
\partial ^{2}}{\partial {\phi }^{2}}}\mathit{\Phi }_{2}+{m}^{2}{r}^{3}\left(
\sin \left( \theta \right) \right) ^{3}f\left( r\right) {\frac{\partial }{%
\partial t}}\mathit{\Phi }_{0}+{m}^{2}{r}^{3}\left( \sin \left( \theta
\right) \right) ^{3}f\left( r\right) {\frac{\partial }{\partial r}}\mathit{%
\Phi }_{1}  \notag \\
& & -{m}^{2}{r}^{3}\left( \sin \left( \theta \right) \right) ^{3}f\left(
r\right) {\frac{\partial }{\partial \phi }}\mathit{\Phi }_{3}+2\,{h}^{2}{r}%
^{3}\left( \sin \left( \theta \right) \right) ^{3}{\frac{\partial ^{3}}{%
\partial {t}^{2}\partial \phi }}\mathit{\Phi }_{3}  \notag \\
& & -4{\hbar }^{2}f\left( r\right) \sin \left( \theta \right) {\frac{%
\partial ^{2}}{\partial {\phi }^{2}}}\mathit{\Phi }_{1}+{m}^{2}{r}^{3}\left( 
{\frac{\partial }{\partial \theta }}\mathit{\Phi }_{2}\right)\left( \sin
\left( \theta \right) \right) ^{3}f\left( r\right)  \notag \\
& & +\left( {\frac{\partial ^{3}}{\partial t\partial {\phi }^{2}}}\mathit{%
\Phi }_{0}\right) f\left( r\right) r\sin \left( \theta \right) {\hbar }%
^{2}+\left( {\frac{\partial ^{3}}{\partial r\partial {\phi }^{2}}}\mathit{%
\Phi }_{1}\right) f\left( r\right) r\sin \left( \theta \right) {\hbar }^{2} 
\notag \\
& & -\left( {\frac{\partial ^{3}}{\partial {\phi }^{3}}}\mathit{\Phi }%
_{3}\right) f\left( r\right) r\sin \left( \theta \right) {\hbar }^{2}-2\,{%
\hbar }^{2}\left( f\left( r\right) \right) ^{2}{r}^{3}\left( \sin \left(
\theta \right) \right) ^{3}{\frac{\partial ^{3}}{\partial {r}^{2}\partial
\phi }}\mathit{\Phi }_{3}-8\,{\hbar }^{2}{r}^{2}\left( \sin \left( \theta
\right) \right) ^{3}\left( f\left( r\right) \right) ^{2}{\frac{\partial ^{2}%
}{\partial r\partial \phi }}\mathit{\Phi }_{3}  \notag \\
& & -4\,{\hbar }^{2}r\left( \sin \left( \theta \right) \right) ^{3}\left( {%
\frac{\partial }{\partial \phi }}\mathit{\Phi }_{3}\right) \left( f\left(
r\right) \right) ^{2}-4\,{\hbar }^{2}f\left( r\right) \left( \sin \left(
\theta \right) \right) ^{2}\left( {\frac{\partial }{\partial \theta }}%
\mathit{\Phi }_{1}\right) \cos \left( \theta \right)  \notag \\
& & -2\,{\hbar }^{2}{r}^{3}\left( \sin \left( \theta \right) \right)
^{3}\left( {\frac{\partial ^{2}}{\partial r\partial \phi }}\mathit{\Phi }%
_{3}\right) f\left( r\right) {\frac{d}{dr}}f\left( r\right) \Bigg]\bigg/%
\bigg[-{\hbar }^{2}f\left( r\right) \sin \left( \theta \right) r\bigg] 
\tag{A6}
\end{align*}

\begin{align*}
\mathcal{G}_{rr}& = & & \Bigg[2\,{\hbar }^{2}f\left( r\right) \left( {\frac{d%
}{dr}}f\left( r\right) \right) r\sin \left( \theta \right) \cos \left(
\theta \right) {\frac{\partial }{\partial \theta }}\mathit{\Phi }_{1}-4\,{%
\hbar }^{2}{r}^{3}\left( \sin \left( \theta \right) \right) ^{2}\left( {%
\frac{d}{dr}}f\left( r\right) \right) f\left( r\right) \mathit{\Phi }_{1}{%
\frac{d^{2}}{d{r}^{2}}}f\left( r\right) \\
& & & +{\hbar }^{2}{r}^{3}\left( \sin \left( \theta \right) \right)
^{2}\left( {\frac{d}{dr}}f\left( r\right) \right) ^{2}\left( {\frac{\partial 
}{\partial t}}\mathit{\Phi }_{0}\right) f\left( r\right) -{\hbar }^{2}{r}%
^{3}\left( \sin \left( \theta \right) \right) ^{2}\left( {\frac{d}{dr}}%
f\left( r\right) \right) ^{2}\left( {\frac{\partial }{\partial r}}\mathit{%
\Phi }_{1}\right) f\left( r\right) \\
& & & -4\,{\hbar }^{2}f\left( r\right) \left( \sin \left( \theta \right)
\right) ^{2}\left( {\frac{d}{dr}}f\left( r\right) \right) ^{2}{r}^{2}\mathit{%
\Phi }_{1}+4\,{\hbar }^{2}\left( f\left( r\right) \right) ^{2}\left( \sin
\left( \theta \right) \right) ^{2}\left( {\frac{d}{dr}}f\left( r\right)
\right) {r}^{2}{\frac{\partial }{\partial r}}\mathit{\Phi }_{1} \\
& & & -8\,{\hbar }^{2}\left( f\left( r\right) \right) ^{2}\left( \sin \left(
\theta \right) \right) ^{2}r\mathit{\Phi }_{1}{\frac{d}{dr}}f\left( r\right)
+8\,{\hbar }^{2}\left( f\left( r\right) \right) ^{3}r\sin \left( \theta
\right) \cos \left( \theta \right) \mathit{\Phi }_{2}+4\,{\hbar }^{2}\left(
f\left( r\right) \right) ^{3} \\
& & & \times {r}^{2}\sin \left( \theta \right) \cos \left( \theta \right) {%
\frac{\partial }{\partial r}}\mathit{\Phi }_{2}+2\,{m}^{2}f\left( r\right) {r%
}^{3}\left( \sin \left( \theta \right) \right) ^{2}\mathit{\Phi }_{1}{\frac{d%
}{dr}}f\left( r\right) +2\,{\hbar }^{2}{r}^{3}\left( \sin \left( \theta
\right) \right) ^{2} \\
& & & \times \left( {\frac{d}{dr}}f\left( r\right) \right) \left( f\left(
r\right) \right) ^{2}{\frac{\partial ^{2}}{\partial t\partial r}}\mathit{%
\Phi }_{0}+4\,{\hbar }^{2}{r}^{3}\left( \sin \left( \theta \right) \right)
^{2}\left( {\frac{d^{2}}{d{r}^{2}}}f\left( r\right) \right) \left( {\frac{%
\partial }{\partial r}}\mathit{\Phi }_{1}\right) \left( f\left( r\right)
\right) ^{2} \\
& & & +2\,{\hbar }^{2}f\left( r\right) \left( \sin \left( \theta \right)
\right) ^{2}\left( {\frac{d}{dr}}f\left( r\right) \right) r{\frac{\partial
^{2}}{\partial {\theta }^{2}}}\mathit{\Phi }_{1}+2\,{\hbar }^{2}{r}%
^{3}\left( \sin \left( \theta \right) \right) ^{2}\left( f\left( r\right)
\right) ^{2}\mathit{\Phi }_{1}{\frac{d^{3}}{d{r}^{3}}}f\left( r\right) \\
& & & +4\,{\hbar }^{2}\left( f\left( r\right) \right) ^{2}\left( \sin \left(
\theta \right) \right) ^{2}{r}^{2}\mathit{\Phi }_{1}{\frac{d^{2}}{d{r}^{2}}}%
f\left( r\right) -2\,{\hbar }^{2}\left( f\left( r\right) \right) ^{2}r\sin
\left( \theta \right) \cos \left( \theta \right) {\frac{\partial ^{2}}{%
\partial \theta \partial r}}\mathit{\Phi }_{1} \\
& & & +4\,{\hbar }^{2}\left( f\left( r\right) \right) ^{2}\left( \sin \left(
\theta \right) \right) ^{2}{\frac{\partial ^{2}}{\partial {\theta }^{2}}}%
\mathit{\Phi }_{1}-2\,{\hbar }^{2}\left( f\left( r\right) \right) ^{2}r{%
\frac{\partial ^{3}}{\partial r\partial {\phi }^{2}}}\mathit{\Phi }_{1}+4\,{%
\hbar }^{2}\left( f\left( r\right) \right) ^{3} \\
& & & \times \left( \sin \left( \theta \right) \right) ^{2}r{\frac{\partial 
}{\partial \theta }}\mathit{\Phi }_{2}+8\,{\hbar }^{2}\left( f\left(
r\right) \right) ^{3}\left( \sin \left( \theta \right) \right) ^{2}r{\frac{%
\partial }{\partial r}}\mathit{\Phi }_{1} \\
& & & +4\,{\hbar }^{2}\left( f\left( r\right) \right) ^{3}r\left( \sin
\left( \theta \right) \right) ^{2}{\frac{\partial }{\partial \phi }}\mathit{%
\Phi }_{3}+4\,{\hbar }^{2}\left( f\left( r\right) \right) ^{2}\sin \left(
\theta \right) \cos \left( \theta \right) {\frac{\partial }{\partial \theta }%
}\mathit{\Phi }_{1} \\
& & & -\left( {\frac{\partial ^{3}}{\partial {r}^{3}}}\mathit{\Phi }%
_{1}\right) \left( f\left( r\right) \right) ^{3}{r}^{3}\left( \sin \left(
\theta \right) \right) ^{2}{\hbar }^{2}+\left( {\frac{\partial ^{3}}{%
\partial t\partial {r}^{2}}}\mathit{\Phi }_{0}\right) \left( f\left(
r\right) \right) ^{3}{r}^{3}\left( \sin \left( \theta \right) \right) ^{2}{%
\hbar }^{2}+ \\
& & & \left( {\frac{\partial ^{3}}{\partial {r}^{2}\partial \phi }}\mathit{%
\Phi }_{3}\right) \left( f\left( r\right) \right) ^{3}{r}^{3}\left( \sin
\left( \theta \right) \right) ^{2}{\hbar }^{2}-4\,{\hbar }^{2}{r}^{3}\left(
\sin \left( \theta \right) \right) ^{2}\left( {\frac{\partial ^{2}}{\partial 
{t}^{2}}}\mathit{\Phi }_{1}\right) {\frac{d}{dr}}f\left( r\right) \\
& & & +4\,{\hbar }^{2}\left( f\left( r\right) \right) ^{3}{r}^{2}\left( \sin
\left( \theta \right) \right) ^{2}{\frac{\partial ^{2}}{\partial r\partial
\phi }}\mathit{\Phi }_{3}+2\,{\hbar }^{2}f\left( r\right) \left( {\frac{d}{dr%
}}f\left( r\right) \right) r{\frac{\partial ^{2}}{\partial {\phi }^{2}}}%
\mathit{\Phi }_{1} \\
& & & -4{\hbar }^{2}\left( f\left( r\right) \right) ^{3}\left( \sin \left(
\theta \right) \right) ^{2}{r}^{2}{\frac{\partial ^{2}}{\partial {r}^{2}}}%
\mathit{\Phi }_{1}+4\,{\hbar }^{2}\left( f\left( r\right) \right) ^{3}\left(
\sin \left( \theta \right) \right) ^{2}{r}^{2}{\frac{\partial ^{2}}{\partial
\theta \partial r}}\mathit{\Phi }_{2} \\
& & & -2\,{\hbar }^{2}\left( f\left( r\right) \right) ^{2}\left( \sin \left(
\theta \right) \right) ^{2}r{\frac{\partial ^{3}}{\partial {\theta }%
^{2}\partial r}}\mathit{\Phi }_{1}+2\,{\hbar }^{2}{r}^{3}\left( \sin \left(
\theta \right) \right) ^{2}\left( {\frac{\partial ^{3}}{\partial {t}%
^{2}\partial r}}\mathit{\Phi }_{1}\right) f\left( r\right) \\
& & & +{\hbar }^{2}{r}^{3}\left( \sin \left( \theta \right) \right) ^{2}%
\mathit{\Phi }_{1}\left( {\frac{d}{dr}}f\left( r\right) \right) ^{3}-{m}%
^{2}\left( f\left( r\right) \right) ^{2}{r}^{3}\left( \sin \left( \theta
\right) \right) ^{2}{\frac{\partial }{\partial r}}\mathit{\Phi }_{1} \\
& & & +{m}^{2}\left( f\left( r\right) \right) ^{2}{r}^{3}\left( \sin \left(
\theta \right) \right) ^{2}{\frac{\partial }{\partial t}}\mathit{\Phi }_{0}+{%
m}^{2}\left( f\left( r\right) \right) ^{2}{r}^{3}\left( \sin \left( \theta
\right) \right) ^{2}{\frac{\partial }{\partial \theta }}\mathit{\Phi }_{2} \\
& & & +{m}^{2}\left( f\left( r\right) \right) ^{2}{r}^{3}\left( \sin \left(
\theta \right) \right) ^{2}{\frac{\partial }{\partial \phi }}\mathit{\Phi }%
_{3}+4{\hbar }^{2}\left( f\left( r\right) \right) ^{2}{\frac{\partial ^{2}}{%
\partial {\phi }^{2}}}\mathit{\Phi }_{1}\Bigg]\bigg/\bigg[-\left( f\left(
r\right) \right) ^{3}{r}^{3}\left( \sin \left( \theta \right) \right) ^{2}{%
\hbar }^{2}\bigg]  \tag{A7}
\end{align*}

\begin{align*}
\mathcal{G}_{\phi r}=\mathcal{G}_{r\phi }& = & & \Bigg[2\,{\hbar }^{2}\left(
\sin \left( \theta \right) \right) ^{4}\left( f\left( r\right) \right) ^{2}{r%
}^{2}{\frac{\partial }{\partial r}}\mathit{\Phi }_{3}-4\,{\hbar }^{2}\left(
f\left( r\right) \right) ^{2}\left( \sin \left( \theta \right) \right) ^{4}r%
\mathit{\Phi }_{3}-2\,{\hbar }^{2}\left( \sin \left( \theta \right) \right)
^{4}\left( f\left( r\right) \right) ^{3}{r}^{2}{\frac{\partial }{\partial r}}%
\mathit{\Phi }_{3} \\
& & & +20\,{\hbar }^{2}\left( f\left( r\right) \right) ^{3}\left( \sin
\left( \theta \right) \right) ^{4}r\mathit{\Phi }_{3}-4\,{m}^{2}{r}%
^{3}\left( \sin \left( \theta \right) \right) ^{4}\left( f\left( r\right)
\right) ^{2}\mathit{\Phi }_{3}-2\,{m}^{2}{r}^{4}\left( \sin \left( \theta
\right) \right) ^{4}\left( f\left( r\right) \right) ^{2}{\frac{\partial }{%
\partial r}}\mathit{\Phi }_{3} \\
& & & -2\,{m}^{2}{r}^{2}\left( \sin \left( \theta \right) \right)
^{2}f\left( r\right) {\frac{\partial }{\partial \phi }}\mathit{\Phi }_{1}-2\,%
{\hbar }^{2}\left( \sin \left( \theta \right) \right) ^{4}{r}^{2}\left(
f\left( r\right) \right) ^{2}{\frac{\partial ^{3}}{\partial {\theta }%
^{2}\partial r}}\mathit{\Phi }_{3}+2\,{\hbar }^{2}{r}^{4}\left( \sin \left(
\theta \right) \right) ^{4} \\
& & & \times \left( {\frac{\partial ^{3}}{\partial {t}^{2}\partial r}}%
\mathit{\Phi }_{3}\right) f\left( r\right) -2\,{\hbar }^{2}{r}^{4}\left(
\sin \left( \theta \right) \right) ^{4}\left( {\frac{d}{dr}}f\left( r\right)
\right) {\frac{\partial ^{2}}{\partial {t}^{2}}}\mathit{\Phi }_{3} \\
& & & -12\,{\hbar }^{2}\left( \sin \left( \theta \right) \right) ^{4}\left(
f\left( r\right) \right) ^{3}{r}^{3}{\frac{\partial ^{2}}{\partial {r}^{2}}}%
\mathit{\Phi }_{3}+4\,{\hbar }^{2}{r}^{3}\left( \sin \left( \theta \right)
\right) ^{4}\left( {\frac{\partial ^{2}}{\partial {t}^{2}}}\mathit{\Phi }%
_{3}\right) f\left( r\right) \\
& & & +2\,\left( {\frac{\partial ^{3}}{\partial t\partial r\partial \phi }}%
\mathit{\Phi }_{0}\right) \left( f\left( r\right) \right) ^{2}{r}^{2}\left(
\sin \left( \theta \right) \right) ^{2}{\hbar }^{2}+2\left( {\frac{\partial
^{3}}{\partial \theta \partial r\partial \phi }}\mathit{\Phi }_{2}\right)
\left( f\left( r\right) \right) ^{2}{r}^{2}\left( \sin \left( \theta \right)
\right) ^{2}{\hbar }^{2} \\
& & & -2\,{\hbar }^{2}f\left( r\right) \cos \left( \theta \right) \sin
\left( \theta \right) {\frac{\partial ^{2}}{\partial \theta \partial \phi }}%
\mathit{\Phi }_{1}+4\,{\hbar }^{2}\left( f\left( r\right) \right)
^{2}r\left( {\frac{\partial ^{2}}{\partial \theta \partial \phi }}\mathit{%
\Phi }_{2}\right) \left( \sin \left( \theta \right) \right) ^{2} \\
& & & +4\,{\hbar }^{2}\left( f\left( r\right) \right) ^{2}\left( \sin \left(
\theta \right) \right) ^{2}r{\frac{\partial ^{2}}{\partial {\phi }^{2}}}%
\mathit{\Phi }_{3}-8\,{\hbar }^{2}\left( \sin \left( \theta \right) \right)
^{2}\left( f\left( r\right) \right) ^{2}{\frac{\partial ^{2}}{\partial
r\partial \phi }}\mathit{\Phi }_{1}r \\
& & & -2\,{\hbar }^{2}\left( \sin \left( \theta \right) \right) ^{4}\left(
f\left( r\right) \right) ^{3}{r}^{4}{\frac{\partial ^{3}}{\partial {r}^{3}}}%
\mathit{\Phi }_{3}-2\,{\hbar }^{2}f\left( r\right) {\frac{\partial ^{3}}{%
\partial {\phi }^{3}}}\mathit{\Phi }_{1}-14{\hbar }^{2}{r}^{3}\left( \sin
\left( \theta \right) \right) ^{4} \\
& & & \times \left( {\frac{d}{dr}}f\left( r\right) \right) \left( f\left(
r\right) \right) ^{2}{\frac{\partial }{\partial r}}\mathit{\Phi }_{3}+10\,{%
\hbar }^{2}r\left( \sin \left( \theta \right) \right) ^{2}\left( {\frac{d}{dr%
}}f\left( r\right) \right) f\left( r\right) {\frac{\partial }{\partial \phi }%
}\mathit{\Phi }_{1}+8\,{\hbar }^{2}\left( f\left( r\right) \right)
^{2}r\left( \sin \left( \theta \right) \right) ^{3}\left( {\frac{\partial }{%
\partial \theta }}\mathit{\Phi }_{3}\right) \\
& & & \times \cos \left( \theta \right) -4\,{\hbar }^{2}\sin \left( \theta
\right) \left( f\left( r\right) \right) ^{2}r\cos \left( \theta \right) {%
\frac{\partial }{\partial \phi }}\mathit{\Phi }_{2}-4\,{\hbar }^{2}\left(
\sin \left( \theta \right) \right) ^{4}\left( f\left( r\right) \right) ^{2}{r%
}^{2}\mathit{\Phi }_{3}{\frac{d}{dr}}f\left( r\right) +16\,{\hbar }%
^{2}r\left( \sin \left( \theta \right) \right) ^{2}\mathit{\Phi }_{3} \\
& & & \times \left( f\left( r\right) \right) ^{2}\left( \cos \left( \theta
\right) \right) ^{2}-{\hbar }^{2}\left( \sin \left( \theta \right) \right)
^{4}\left( f\left( r\right) \right) ^{2}{r}^{4}\left( {\frac{\partial }{%
\partial r}}\mathit{\Phi }_{3}\right) {\frac{d^{2}}{d{r}^{2}}}f\left(
r\right) +{\hbar }^{2}\left( \sin \left( \theta \right) \right) ^{2}f\left(
r\right) {r}^{2}\left( {\frac{d^{2}}{d{r}^{2}}}f\left( r\right) \right) {%
\frac{\partial }{\partial \phi }}\mathit{\Phi }_{1} \\
& & & -4\,{\hbar }^{2}{r}^{4}\left( \sin \left( \theta \right) \right)
^{4}\left( {\frac{d}{dr}}f\left( r\right) \right) \left( f\left( r\right)
\right) ^{2}{\frac{\partial ^{2}}{\partial {r}^{2}}}\mathit{\Phi }_{3}-2\,{%
\hbar }^{2}\left( \sin \left( \theta \right) \right) ^{4}\left( f\left(
r\right) \right) ^{2}{r}^{3}\mathit{\Phi }_{3}{\frac{d^{2}}{d{r}^{2}}}%
f\left( r\right) -6\,{\hbar }^{2}\left( \sin \left( \theta \right) \right)
^{3}{r}^{2} \\
& & & \times \left( f\left( r\right) \right) ^{2}\left( {\frac{\partial ^{2}%
}{\partial \theta \partial r}}\mathit{\Phi }_{3}\right) \cos \left( \theta
\right) -4\,{\hbar }^{2}\sin \left( \theta \right) {r}^{2}\cos \left( \theta
\right) \left( f\left( r\right) \right) ^{2}{\frac{\partial ^{2}}{\partial
r\partial \phi }}\mathit{\Phi }_{2}+2{\hbar }^{2}{r}^{2}\left( \sin \left(
\theta \right) \right) ^{2}\left( {\frac{d}{dr}}f\left( r\right) \right) \\
& & & \times \left( {\frac{\partial ^{2}}{\partial t\partial \phi }}\mathit{%
\Phi }_{0}\right) f\left( r\right) -2\,{\hbar }^{2}\left( \sin \left( \theta
\right) \right) ^{2}f\left( r\right) {\frac{\partial }{\partial \phi }}%
\mathit{\Phi }_{1}+10\,{\hbar }^{2}\left( \sin \left( \theta \right) \right)
^{2}\left( f\left( r\right) \right) ^{2}{\frac{\partial }{\partial \phi }}%
\mathit{\Phi }_{1} \\
& & & -2\,{\hbar }^{2}f\left( r\right) \left( \sin \left( \theta \right)
\right) ^{2}{\frac{\partial ^{3}}{\partial {\theta }^{2}\partial \phi }}%
\mathit{\Phi }_{1}+2{\hbar }^{2}{r}^{2}\left( \sin \left( \theta \right)
\right) ^{2}{\frac{\partial ^{3}}{\partial {t}^{2}\partial \phi }}\mathit{%
\Phi }_{1}\Bigg]\bigg/\bigg[-2{\hbar }^{2}\left( \sin \left( \theta \right)
\right) ^{2}\left( f\left( r\right) \right) ^{2}{r}^{2}\bigg]  \tag{A8}
\end{align*}

\begin{align*}
\mathcal{G}_{\phi \theta }=\mathcal{G}_{\theta \phi} &= & \Bigg[-{\hbar }%
^{2}f\left( r\right) r{\frac{\partial ^{3}}{\partial {\phi }^{3}}}\mathit{%
\Phi }_{2}+{\hbar }^{2}{r}^{3}\left( \sin \left( \theta \right) \right) ^{2}{%
\frac{\partial ^{3}}{\partial {t}^{2}\partial \phi }}\mathit{\Phi }_{2}+{%
\hbar }^{2}{r}^{3}\left( \sin \left( \theta \right) \right) ^{4}{\frac{%
\partial ^{3}}{\partial \theta \partial {t}^{2}}}\mathit{\Phi }_{3} \\
& & -4\,{\hbar }^{2}f\left( r\right) \left( \sin \left( \theta \right)
\right) ^{2}{\frac{\partial ^{2}}{\partial \theta \partial \phi }}\mathit{%
\Phi }_{1}-4\,{\hbar }^{2}{r}^{2}\left( \sin \left( \theta \right) \right)
^{2}\left( f\left( r\right) \right) ^{2}{\frac{\partial ^{2}}{\partial
r\partial \phi }}\mathit{\Phi }_{2} \\
& & -4{\hbar }^{2}{r}^{2}\left( \sin \left( \theta \right) \right)
^{4}\left( {\frac{\partial ^{2}}{\partial \theta \partial r}}\mathit{\Phi }%
_{3}\right) \left( f\left( r\right) \right) ^{2} \\
& & +2\,{\hbar }^{2}{r}^{3}\left( \sin \left( \theta \right) \right)
^{3}\left( {\frac{\partial ^{2}}{\partial {t}^{2}}}\mathit{\Phi }_{3}\right)
\cos \left( \theta \right) +5\,{\hbar }^{2}r\left( \sin \left( \theta
\right) \right) ^{4}\left( {\frac{\partial }{\partial \theta }}\mathit{\Phi }%
_{3}\right) f\left( r\right) \\
& & -2\,{\hbar }^{2}r\left( \sin \left( \theta \right) \right) ^{4}\left( {%
\frac{\partial }{\partial \theta }}\mathit{\Phi }_{3}\right) \left( f\left(
r\right) \right) ^{2}-2\,{\hbar }^{2}\left( f\left( r\right) \right)
^{2}r\left( \sin \left( \theta \right) \right) ^{2}  \notag \\
& & \times {\frac{\partial }{\partial \phi }}\mathit{\Phi }_{2}+4\,{\hbar }%
^{2}f\left( r\right) \sin \left( \theta \right) \cos \left( \theta \right) {%
\frac{\partial }{\partial \phi }}\mathit{\Phi }_{1}+3\,{\hbar }^{2}f\left(
r\right) r\left( \cos \left( \theta \right) \right) ^{2}{\frac{\partial }{%
\partial \phi }}\mathit{\Phi }_{2} \\
& & -{\hbar }^{2}f\left( r\right) r\left( \sin \left( \theta \right) \right)
^{4}{\frac{\partial ^{3}}{\partial {\theta }^{3}}}\mathit{\Phi }_{3}-{\hbar }%
^{2}{r}^{3}\left( \sin \left( \theta \right) \right) ^{2}\left( f\left(
r\right) \right) ^{2}{\frac{\partial ^{3}}{\partial {r}^{2}\partial \phi }}%
\mathit{\Phi }_{2} \\
& & -{\hbar }^{2}{r}^{3}\left( \sin \left( \theta \right) \right) ^{4}\left(
f\left( r\right) \right) ^{2}{\frac{\partial ^{3}}{\partial \theta \partial {%
r}^{2}}}\mathit{\Phi }_{3}+\left( {\frac{\partial ^{3}}{\partial \theta
\partial r\partial \phi }}\mathit{\Phi }_{1}\right) f\left( r\right) r\left(
\sin \left( \theta \right) \right) ^{2}{\hbar }^{2}  \notag \\
& & +\left( {\frac{\partial ^{3}}{\partial \theta \partial t\partial \phi }}%
\mathit{\Phi }_{0}\right) f\left( r\right) r\left( \sin \left( \theta
\right) \right) ^{2}{\hbar }^{2}-{\hbar }^{2}f\left( r\right) r\left( \sin
\left( \theta \right) \right) ^{2}{\frac{\partial }{\partial \phi }}\mathit{%
\Phi }_{2} \\
& & -{m}^{2}f\left( r\right) {r}^{3}\left( \sin \left( \theta \right)
\right) ^{4}{\frac{\partial }{\partial \theta }}\mathit{\Phi }_{3}-{m}%
^{2}f\left( r\right) {r}^{3}\left( \sin \left( \theta \right) \right) ^{2}{%
\frac{\partial }{\partial \phi }}\mathit{\Phi }_{2} \\
& & +4{\hbar }^{2}r\left( \sin \left( \theta \right) \right) ^{3}\mathit{%
\Phi }_{3}f\left( r\right) \cos \left( \theta \right) \\
& & -{\hbar }^{2}{r}^{3}\left( \sin \left( \theta \right) \right)
^{4}f\left( r\right) \times \left( {\frac{d}{dr}}f\left( r\right) \right) {%
\frac{\partial ^{2}}{\partial \theta \partial r}}\mathit{\Phi }_{3}-{\hbar }%
^{2}{r}^{3}\left( \sin \left( \theta \right) \right) ^{2}  \notag \\
& & \times f\left( r\right) \left( {\frac{d}{dr}}f\left( r\right) \right) {%
\frac{\partial ^{2}}{\partial r\partial \phi }}\mathit{\Phi }_{2}+{\hbar }%
^{2}f\left( r\right) r\left( \sin \left( \theta \right) \right) ^{2}\left( {%
\frac{\partial }{\partial \theta }}\mathit{\Phi }_{3}\right) \left( \cos
\left( \theta \right) \right) ^{2} \\
& & -2\,{\hbar }^{2}{r}^{3}\left( \sin \left( \theta \right) \right)
^{3}\left( f\left( r\right) \right) ^{2}\times \left( {\frac{\partial ^{2}}{%
\partial {r}^{2}}}\mathit{\Phi }_{3}\right)\cos \left( \theta \right) \\
& & -5{\hbar }^{2}f\left( r\right) r\left( \sin \left( \theta \right)
\right) ^{3}\left( {\frac{\partial ^{2}}{\partial {\theta }^{2}}}\mathit{%
\Phi }_{3}\right) \cos \left( \theta \right) -3\,{\hbar }^{2}f\left(
r\right) \cos \left( \theta \right) r\left( {\frac{\partial ^{2}}{\partial
\theta \partial \phi }}\mathit{\Phi }_{2}\right) \sin \left( \theta \right)
\\
& & +2{\hbar }^{2}f\left( r\right) r\sin \left( \theta \right) \left( {\frac{%
\partial ^{2}}{\partial {\phi }^{2}}}\mathit{\Phi }_{3}\right) \cos \left(
\theta \right) +4\,{\hbar }^{2}r\left( \sin \left( \theta \right) \right)
^{3}\mathit{\Phi }_{3}\left( f\left( r\right) \right) ^{2}\cos \left( \theta
\right) \\
& & -4\,{\hbar }^{2}{r}^{2}\left( \sin \left( \theta \right) \right)
^{3}\left( {\frac{\partial }{\partial r}}\mathit{\Phi }_{3}\right) \left(
f\left( r\right) \right) ^{2}\cos \left( \theta \right) +8{\hbar }%
^{2}f\left( r\right) \sin \left( \theta \right) r\left( \cos \left( \theta
\right) \right) ^{3}\mathit{\Phi }_{3} \\
& & -2\,{m}^{2}f\left( r\right) {r}^{3}\left( \sin \left( \theta \right)
\right) ^{3}\mathit{\Phi }_{3}\cos \left( \theta \right)-2\,{\hbar }^{2}{r}%
^{3}\left( \sin \left( \theta \right) \right) ^{3}f\left( r\right) \\
& & \times\left( {\frac{d}{dr}}f\left( r\right) \right) \left( {\frac{%
\partial }{\partial r}}\mathit{\Phi }_{3}\right)\cos \left( \theta \right)%
\Bigg] \bigg/\bigg[-{\hbar }^{2}f\left( r\right) \left( \sin \left( \theta
\right) \right) ^{2}r\bigg]  \tag{A9}
\end{align*}

\begin{align*}
\mathcal{G}_{tr}=\mathcal{G}_{rt}& = & & \Bigg[-{\hbar }^{2}{r}^{3}\left(
\sin \left( \theta \right) \right) ^{2}{\frac{\partial ^{3}}{\partial {t}^{3}%
}}\mathit{\Phi }_{1}^{2}{\hbar }\left( f\left( r\right) \right) ^{2}\left(
\sin \left( \theta \right) \right) ^{2}{r}^{2}{\frac{\partial ^{2}}{\partial
\theta \partial t}}\mathit{\Phi }_{2} \\
& & & +2\,{\hbar }^{2}{r}^{3}\left( \sin \left( \theta \right) \right)
^{2}\left( {\frac{\partial }{\partial t}}\mathit{\Phi }_{1}\right) \left( {%
\frac{d}{dr}}f\left( r\right) \right) ^{2}+2\,{\hbar }^{2}\left( f\left(
r\right) \right) ^{2}\left( \sin \left( \theta \right) \right) ^{2} \\
& & & \times \left( {\frac{\partial }{\partial t}}\mathit{\Phi }_{1}\right)
r-2\,{\hbar }^{2}\left( f\left( r\right) \right) ^{4}\left( \sin \left(
\theta \right) \right) ^{2}\left( {\frac{\partial }{\partial r}}\mathit{\Phi 
}_{0}\right) r- \\
& & & 2\,{\hbar }^{2}\sin \left( \theta \right) \cos \left( \theta \right)
\left( f\left( r\right) \right) ^{3}{\frac{\partial }{\partial \theta }}%
\mathit{\Phi }_{0}-2\,{\hbar }^{2}\left( f\left( r\right) \right) ^{2}\left(
\sin \left( \theta \right) \right) ^{2}{r}^{2}{\frac{\partial ^{2}}{\partial
t\partial r}}\mathit{\Phi }_{1} \\
& & & +{\hbar }^{2}\left( f\left( r\right) \right) ^{3}\left( \sin \left(
\theta \right) \right) ^{2}\left( {\frac{\partial ^{3}}{\partial {\theta }%
^{2}\partial r}}\mathit{\Phi }_{0}\right) r-{\hbar }^{2}f\left( r\right)
\left( \sin \left( \theta \right) \right) ^{2}r{\frac{\partial ^{3}}{%
\partial {\theta }^{2}\partial t}}\mathit{\Phi }_{1} \\
& & & +{\hbar }^{2}{r}^{3}\left( \sin \left( \theta \right) \right)
^{2}\left( f\left( r\right) \right) ^{4}{\frac{\partial ^{3}}{\partial {r}%
^{3}}}\mathit{\Phi }_{0}+\left( {\frac{\partial ^{3}}{\partial t\partial
r\partial \phi }}\mathit{\Phi }_{3}\right) \\
& & & \times \left( f\left( r\right) \right) ^{2}{r}^{3}\left( \sin \left(
\theta \right) \right) ^{2}{\hbar }^{2}+\left( {\frac{\partial ^{3}}{%
\partial \theta \partial t\partial r}}\mathit{\Phi }_{2}\right) \left(
f\left( r\right) \right) ^{2}{r}^{3}\left( \sin \left( \theta \right)
\right) ^{2}{\hbar }^{2} \\
& & & +{\hbar }^{2}\left( {\frac{d}{dr}}f\left( r\right) \right) \left( {%
\frac{\partial ^{2}}{\partial {\phi }^{2}}}\mathit{\Phi }_{1}\right) \left(
f\left( r\right) \right) ^{2}r-{m}^{2}f\left( r\right) {r}^{3}\left( \sin
\left( \theta \right) \right) ^{2}{\frac{\partial }{\partial t}}\mathit{\Phi 
}_{1} \\
& & & +{m}^{2}\left( f\left( r\right) \right) ^{3}{r}^{3}\left( \sin \left(
\theta \right) \right) ^{2}{\frac{\partial }{\partial r}}\mathit{\Phi }%
_{0}+2\,{\hbar }^{2}{r}^{2}\left( \sin \left( \theta \right) \right)
^{2}\left( {\frac{\partial ^{2}}{\partial t\partial \phi }}\mathit{\Phi }%
_{3}\right) \left( f\left( r\right) \right) ^{2} \\
& & & +2\,{\hbar }^{2}\left( f\left( r\right) \right) ^{4}\left( \sin \left(
\theta \right) \right) ^{2}{r}^{2}{\frac{\partial ^{2}}{\partial {r}^{2}}}%
\mathit{\Phi }_{0}+{\hbar }^{2}{r}^{3}\left( \sin \left( \theta \right)
\right) ^{2}\left( {\frac{d^{3}}{d{r}^{3}}}f\left( r\right) \right) \mathit{%
\Phi }_{0}\left( f\left( r\right) \right) ^{3} \\
& & & +{\hbar }^{2}{r}^{3}\left( \sin \left( \theta \right) \right)
^{2}\left( {\frac{d}{dr}}f\left( r\right) \right) f\left( r\right) {\frac{%
\partial ^{2}}{\partial {t}^{2}}}\mathit{\Phi }_{0}-{\hbar }^{2}{r}%
^{3}\left( \sin \left( \theta \right) \right) ^{2}\left( {\frac{d}{dr}}%
f\left( r\right) \right) \\
& & & \times \left( {\frac{\partial ^{2}}{\partial t\partial r}}\mathit{\Phi 
}_{1}\right) f\left( r\right) +{h}^{2}r\sin \left( \theta \right) \cos
\left( \theta \right) \left( f\left( r\right) \right) ^{3}\left( {\frac{%
\partial ^{2}}{\partial \theta \partial r}}\mathit{\Phi }_{0}\right) \\
& & & -{\hbar }^{2}f\left( r\right) \sin \left( \theta \right) \cos \left(
\theta \right) r{\frac{\partial ^{2}}{\partial \theta \partial t}}\mathit{%
\Phi }_{1}-{\hbar }^{2}{r}^{3}f\left( r\right) \left( \sin \left( \theta
\right) \right) ^{2}\left( {\frac{d}{dr}}f\left( r\right) \right) ^{3}%
\mathit{\Phi }_{0} \\
& & & +{\hbar }^{2}{r}^{3}\left( \sin \left( \theta \right) \right)
^{2}\left( {\frac{d}{dr}}f\left( r\right) \right) ^{2}\left( f\left(
r\right) \right) ^{2}{\frac{\partial }{\partial r}}\mathit{\Phi }_{0}+{m}%
^{2}\left( f\left( r\right) \right) ^{2}{r}^{3}\left( \sin \left( \theta
\right) \right) ^{2}\left( {\frac{d}{dr}}f\left( r\right) \right) \\
& & & \times \mathit{\Phi }_{0}+{h}^{2}\left( f\left( r\right) \right)
^{2}\left( \sin \left( \theta \right) \right) ^{2}\left( {\frac{d}{dr}}%
f\left( r\right) \right) \left( {\frac{\partial ^{2}}{\partial {\theta }^{2}}%
}\mathit{\Phi }_{0}\right) r \\
& & & +{h}^{2}{r}^{3}\left( \sin \left( \theta \right) \right) ^{2}f\left(
r\right) \left( {\frac{\partial }{\partial t}}\mathit{\Phi }_{1}\right) {%
\frac{d^{2}}{d{r}^{2}}}f\left( r\right) +4\,{h}^{2}{r}^{3}\left( \sin \left(
\theta \right) \right) ^{2}\left( {\frac{d}{dr}}f\left( r\right) \right)
\left( f\left( r\right) \right) ^{3} \\
& & & \times {\frac{\partial ^{2}}{\partial {r}^{2}}}\mathit{\Phi }_{0}+2\,{%
\hbar }^{2}\left( f\left( r\right) \right) ^{3}\left( \sin \left( \theta
\right) \right) ^{2}{r}^{2}\left( {\frac{d^{2}}{d{r}^{2}}}f\left( r\right)
\right) \mathit{\Phi }_{0} \\
& & & +4\,{\hbar }^{2}\left( f\left( r\right) \right) ^{3}\left( \sin \left(
\theta \right) \right) ^{2}{r}^{2}\left( {\frac{d}{dr}}f\left( r\right)
\right) {\frac{\partial }{\partial r}}\mathit{\Phi }_{0}+2\,{\hbar }%
^{2}f\left( r\right) \left( \sin \left( \theta \right) \right) ^{2}\left( {%
\frac{d}{dr}}f\left( r\right) \right) {r}^{2} \\
& & & \times {\frac{\partial }{\partial t}}\mathit{\Phi }_{1}-2\,{\hbar }%
^{2}\left( f\left( r\right) \right) ^{3}\left( \sin \left( \theta \right)
\right) ^{2}\left( {\frac{d}{dr}}f\left( r\right) \right) \mathit{\Phi }_{0}r
\\
& & & +2\,{\hbar }^{2}\sin \left( \theta \right) \cos \left( \theta \right) {%
r}^{2}\left( {\frac{\partial }{\partial t}}\mathit{\Phi }_{2}\right) \left(
f\left( r\right) \right) ^{2}+3\,{\hbar }^{2}{r}^{3}\left( \sin \left(
\theta \right) \right) ^{2}\left( {\frac{d^{2}}{d{r}^{2}}}f\left( r\right)
\right) \\
& & & \times \left( f\left( r\right) \right) ^{3}{\frac{\partial }{\partial r%
}}\mathit{\Phi }_{0}-2\,{\hbar }^{2}\left( f\left( r\right) \right) ^{3}{%
\frac{\partial ^{2}}{\partial {\phi }^{2}}}\mathit{\Phi }_{0} \\
& & & +{\hbar }^{2}{r}^{3}\left( \sin \left( \theta \right) \right)
^{2}\left( {\frac{d}{dr}}f\left( r\right) \right) \left( {\frac{d^{2}}{d{r}%
^{2}}}f\left( r\right) \right) \mathit{\Phi }_{0}\left( f\left( r\right)
\right) ^{2} \\
& & & +{\hbar }^{2}\sin \left( \theta \right) \cos \left( \theta \right)
\left( {\frac{d}{dr}}f\left( r\right) \right) \left( {\frac{\partial }{%
\partial \theta }}\mathit{\Phi }_{0}\right) \left( f\left( r\right) \right)
^{2}r \\
& & & +{\hbar }^{2}\left( f\left( r\right) \right) ^{3}\left( {\frac{%
\partial ^{3}}{\partial r\partial {\phi }^{2}}}\mathit{\Phi }_{0}\right) r-{%
\hbar }^{2}f\left( r\right) r{\frac{\partial ^{3}}{\partial t\partial {\phi }%
^{2}}}\mathit{\Phi }_{1} \\
& & & -2\,{\hbar }^{2}\left( f\left( r\right) \right) ^{3}\left( \sin \left(
\theta \right) \right) ^{2}{\frac{\partial ^{2}}{\partial {\theta }^{2}}}%
\mathit{\Phi }_{0}\Bigg]\bigg/\bigg[\left( f\left( r\right) \right) ^{2}{r}%
^{3}\left( \sin \left( \theta \right) \right) ^{2}\hbar ^{2}\bigg]  \tag{A10}
\end{align*}

\end{document}